\documentclass[%
 reprint,
 amsmath,amssymb,
aps,prl,onecolumn, superscriptaddress, showpacs
]{revtex4-1}

\def \bra#1{\mathinner{\langle{#1|}}}
\def \ket#1{\mathinner{|{#1}\rangle}}

\def \red #1 {\textcolor{red}{#1}}

\usepackage{graphicx}

\usepackage{dcolumn}
\usepackage{bm}
\usepackage{mathtools}
\usepackage{xcolor}
\usepackage{framed}

\usepackage{setspace} 
\usepackage{algorithm}
\usepackage{algpseudocode}
\floatname{algorithm}{Strategy}

\usepackage{babel}

\usepackage{tikz}
\usetikzlibrary{automata, arrows}
\usepackage{circuitikz}
\usetikzlibrary{backgrounds,fit,decorations.pathreplacing,calc,backgrounds}  
\tikzstyle{operator} = [draw,fill=white,minimum size=1.5em] 
    \tikzstyle{operator2}=[draw,fill=white, text width=0.6cm, minimum height=4cm] 
    \tikzstyle{operator3}=[draw,fill=white, text width=0.8cm, minimum height=1cm] 
    \tikzstyle{phase} = [draw, fill=white,shape=circle,minimum size=12pt,inner sep=0pt]
    \tikzstyle{phase1} = [fill=black, shape=circle,minimum size=3.5pt,inner sep=0pt]
    \tikzstyle{phase2} = [fill=blue, shape=circle,minimum size=7pt,inner sep=1pt]
    \tikzstyle{phase11} = [fill=cyan, shape=circle,minimum size=7pt,inner sep=1pt]
    \tikzstyle{phase12} = [fill=lime, shape=circle,minimum size=7pt,inner sep=1pt]
    \tikzstyle{phase13} = [fill=pink, shape=circle,minimum size=7pt,inner sep=1pt]
    \tikzstyle{phase14} = [fill=lightgray, shape=circle,minimum size=7pt,inner sep=1pt]
    \tikzstyle{phase0} = [draw, fill=white, shape=circle,minimum size=3.5pt,inner sep=0pt]
    \tikzstyle{ellipsis} = [fill,shape=circle,minimum size=2pt,inner sep=0pt]
    \tikzset{meter/.append style={fill=white, draw, inner sep=5, rectangle, font=\vphantom{A}, minimum width=15, 
 path picture={\draw[black] ([shift={(.05,.2)}]path picture bounding box.south west) to[bend left=40] ([shift={(-.05,.2)}]path picture bounding box.south east);\draw[black,-latex] ([shift={(0,.15)}]path picture bounding box.south) -- ([shift={(.15,-.08)}]path picture bounding box.north);}}}
 \tikzset{cross/.style={path picture={\draw[thick,black](path picture bounding box.north) -- (path picture bounding box.south) (path picture bounding box.west) -- (path picture bounding box.east);
   }},
   crossx/.style={path picture={\draw[thick,black,inner sep=0pt]
   (path picture bounding box.south east) -- (path picture bounding box.north west) (path picture bounding box.south west) -- (path picture bounding box.north east);
   }},
   circlewc/.style={draw,circle,cross,minimum width=0.3 cm},
   }

\newcommand*{\Scale}[2][4]{\scalebox{#1}{$#2$}}%

\usepackage[hang,small,bf]{caption}    

\usepackage{tikz} 
\usetikzlibrary{backgrounds,fit,decorations.pathreplacing,calc,backgrounds}  

\begin{document}


\title{Towards a NISQ Algorithm to Simulate Hermitian Matrix Exponentiation}

\author{Keren Li}
\email{likeren1021@gmail.com}
\affiliation{Shenzhen JL Computational Science and Applied Research Institute, Shenzhen 518109, China}

\date{\today}

\begin{abstract}
   A practical fault-tolerant quantum computer is worth looking forward to as it provides applications that outperform their known classical counterparts.   
   However, millions of interacting qubits with stringent criteria are required, which is intractable with current quantum technologies.
   As it would take decades to make it happen, exploiting the power of noisy intermediate-scale quantum(NISQ) devices, which already exist, is becoming one of current goals. 
   Matrix exponentiation, especially hermitian matrix exponentiation, is an essential element for quantum information processing. In this article, a heuristic method is reported as simulating a hermitian matrix exponentiation using parametrized quantum circuit(PQC). To generate PQCs for simulations, two strategies, each with its own advantages, are proposed, and can be deployed on near future quantum devices.
   Compared with the method such as product formula and density matrix exponentiation, the PQCs provided in our method require only low depth circuit and easily accessible gates, which benefit experimental realizations.
   Furthermore, in this paper, an ancilla-assisted parameterized quantum circuit is proposed to characterize and compress a unitary process, which is likely to be applicable to realizing applications on NISQ hardwares, such as phase estimation, principal component analyses, and matrix inversion.
   To support the feasibility of our method, numerical experiments were investigated via simulating evolutions by Bell state, GHZ state and Hamiltonian of Crotonic acid, which show an experimental friendly result when compared with their conventional methods.
   As pursuing a fault-tolerant quantum computer is still challenging and takes decades, our work, which gives a NISQ device friendly way, contributes to the field of NISQ algorithms and provides a possibility, exploiting the power with current quantum technology.
\end{abstract}

\maketitle


\section{introduction}   

As dramatically improved productivities are potentially provided with applying quantum algorithms on cryptography, search, and machine learning, building up a universal quantum computer will be one of the greatest scientific and engineering achievements\cite{shor1999polynomial,grover1997quantum,boixo2014evidence,2017-Seth-Nat-quantum}. 
With rapidly emerging experimental advances, quantum advantages have been demonstrated for sampling the output distributions of either random quantum circuits or identical bosons scattered by a linear interferometer\cite{2019-google-Qsupremacy,zhong2020quantum}. However, the current platforms are still restricted on an order of a hundred physical qubits, which are refereed to as Noisy Intermediate-Scale Quantum(NISQ) devices\cite{2020-honeywell-ion,preskill2018quantum}. For instance, 53-qubit Sycamore chip is such a state of art device, with single-qubit gate fidelities of $99.85\%$ and two-qubit gate fidelities of $99.64\%$, representing the current machines which have accurate basic quantum gates but limited qubits and coherence time\cite{2019-google-Qsupremacy}.

As a fault-tolerant quantum computer with millions of physical qubits is still a daunting goal, one present-day goal is to exploit the power of current quantum hardwares while developing techniques.
Additionally, owing to the stringent requirements to realize conventional quantum algorithms, NISQ algorithms are designed, which have no explicit requirements for qubits with error correction and can thus be deployed on the near term quantum hardwares\cite{preskill2018quantum, bharti2021noisy}.


Matrix exponentiation is an essential element for quantum computers as it provides the capability to execute matrix multiplications in an ultra-fast way. Compared with the classical computation complexity for calculating a $d\times d$ matrix exponentiation being $\mathcal{O}(d^3)$, the internal Hamiltonian exponentiation on a quantum system requires only $\mathcal{O}(1)$. Furthermore, for other kind of matrices exponentiation, density matrix exponentiation and Hamiltonian simulations are provided if matrices are hermitian\cite{lloyd2014quantum,low2019hamiltonian,childs2021theory}. However, implementing those protocols require deep quantum circuit, which hinder their applications on near term devices.

In this paper, a NISQ device friendly framework is proposed to simulate matrix exponentiation, which can be employed in applications such as quantum principal component analysis, hamiltonian simulation, quantum matrix inversion, and their generalizations\cite{lloyd2014quantum,rebentrost2019quantum,gao2021quantum,harrow2009quantum,rebentrost2014quantum}. 
The framework is heuristic and is iteratively implemented with a parametrized quantum circuit(PQC), which can be addressed by two strategies. 
One based on the optimal control theory, directly finds the target PQC but is in general inefficient. The second, averting the inefficiency, generates the target PQC with a hybrid quantum-classical paradigm\cite{mcclean2016theory}. The cost of both strategies are analyzed thereafter. 
Furthermore, numerical experiments are conducted to support the feasibility of the algorithm via investigations on simulating the evolutions steered by Bell state, GHZ state, and Crotonic acid molecular\cite{li2021optimizing}. 
Compared with the method such as product formula and density matrix exponentiation, our method appears more friendly to the experimental realizations.
In addition, an ancilla-assisted parameterized quantum circuit(APQC), as a generalization of linear combination unitaries(LCU) and parameterized quantum circuit\cite{childs2012hamiltonian,long2008duality}, is proposed to generate PQCs for simulations. Though APQC is employed as either characterizing or compressing a quantum circuits in our strategies, it is potential to be an important module for variant quantum algorithms, which are important constitutions for NISQ algorithms.

\section{the framework}
\begin{figure}[!h]
   \centerline{
   \begin{tikzpicture}[thick]
   \ctikzset{scale=1.2}
   \tikzstyle{every node}=[font=\small,scale=0.9]
     \tikzstyle{operator} = [draw,fill=white,minimum size=1em] 
     \tikzstyle{operator2} = [draw,fill=white,minimum size=2em] 
     \tikzstyle{operator3} = [draw,fill=white,minimum size=4em] 
     \tikzstyle{operator4} = [draw,shape=rectangle, fill=white, minimum width=1em, minimum height=10em] 
     \tikzstyle{phase} = [fill,shape=circle,minimum size=3pt,inner sep=0pt]
     \tikzstyle{surround} = [fill=blue!10,thick,draw=black,rounded corners=2mm]
     \tikzstyle{ellipsis} = [fill,shape=circle,minimum size=2pt,inner sep=0pt]
     \tikzstyle{ellipsis2} = [fill,shape=circle,minimum size=0.5pt,inner sep=0pt]
     \node at (0,-1) (q3) {$(a)$};
     \node at (.2,-1.6) (q3) {$\sigma$};
     \node (end3) at (2,-1.6) {} edge [-] (q3);
     \node at (.5,-1.6) (hy111) {$/$};
     \node[operator2] (op22) at (1.2,-1.6) {$e^{-i\rho t}$} ;
     \node at (2.35,-1.6) (eq1) {$\Leftarrow$};
     \node at (3,-1.6) (q3) {$\sigma$};
     \node (end3) at (8.5,-1.6) {} edge [-] (q3);
     \node at (3.3,-1.6) (hy111) {$/$};
     \node[operator2] (op22) at (4.5,-1.6) {$\prod_i e^{-i\rho_i t}$} ;
     \node[operator2] (op22) at (6,-1.6) {$\prod_i e^{-i\rho_i t}$} ;
     \node[operator2] (op22) at (7.5,-1.6) {$\prod_i e^{-i\rho_i t}$} ;
     \node[ellipsis] (ee1) at (8.65,-1.6) {};
     \node[ellipsis] (ee2) at (8.75,-1.6) {};
     \node[ellipsis] (ee3) at (8.85,-1.6) {}; 
     \node at (9,-1.6) (q3) {};
     \node (end3) at (11,-1.6) {} edge [-] (q3);
     \node[operator2] (op22) at (10,-1.6) {$\prod_i e^{-i\rho_i t}$} ;
     \node at (0,-2.5) (q3) {$(b)$};
     \node at (0.2,-3) (q33) {$\sigma$};
     \node at (0.2,-3.6) (q3232) {$\rho$};
     \node at (0.2,-4.2) (q332) {$\rho$};
     \node at (0.2,-4.8) (q333) {$\rho$};
     \node at (0.2,-5.8) (q44) {$\rho$};
     \node at (0.2,-6.4) (q45) {$\rho$};
     \node[ellipsis] (ee1) at (0.2,-5.2) {};
     \node[ellipsis] (ee2) at (0.2,-5.3) {};
     \node[ellipsis] (ee3) at (0.2,-5.4) {};     
     \node at (0.5,-3) (hy33) {$/$};
     \node at (0.5,-3.6) (hy3232) {$/$};
     \node at (0.5,-4.2) (hy332) {$/$};
     \node at (0.5,-4.8) (hy44) {$/$};
     \node at (0.5,-5.8) (hy44) {$/$};
     \node at (0.5,-6.4) (hy44) {$/$};
     \node (end3) at (6.9,-3) {} edge [-] (q33);
     \node (end32) at (6.9,-3.6) {} edge [-] (q3232);      
     \node (end32) at (6.9,-4.2) {} edge [-] (q332);
     \node (end4) at (6.9,-4.8) {} edge [-] (q333); 
     \node (end4) at (6.9,-5.8) {} edge [-] (q44); 
     \node (end4) at (6.9,-6.4) {} edge [-] (q45); 
     \node[operator3] (op21) at (1.3,-3.3) {$e^{-i S \Delta t}$} ;
     \node[ellipsis] (ee1) at (7.1,-4.8) {};
     \node[ellipsis] (ee2) at (7.2,-4.8) {};
     \node[ellipsis] (ee3) at (7.3,-4.8) {}; 
     \node at (2.3,-3.6) (cr122) {$\times$};
     \node at (2.3,-4.2) (cr212) {$\times$};
     \node[ellipsis2] at (2.3,-3.6) (ep122) {};
     \node[ellipsis2] at (2.3,-4.2) (ep212) {};
     \draw[-] (ep122) -- (ep212);
     \node[operator3] (op21) at (3.3,-3.3) {$e^{-i S \Delta t}$} ;
     \node at (4.3,-4.2) (cr121) {$\times$};
     \node at (4.3,-4.8) (cr211) {$\times$};
     \node[ellipsis2] at (4.3,-4.2) (ep121) {};
     \node[ellipsis2] at (4.3,-4.8) (ep211) {};
     \draw[-] (ep121) -- (ep211);
     \node at (4.8,-3.6) (cr121) {$\times$};
     \node at (4.8,-4.2) (cr211) {$\times$};
     \node[ellipsis2] at (4.8,-3.6) (ep121) {};
     \node[ellipsis2] at (4.8,-4.2) (ep211) {};
     \draw[-] (ep121) -- (ep211);
     \node[operator3] (op21) at (5.8,-3.3) {$e^{-i S \Delta t}$} ;
     \node at (7.5,-3)  (q33){};
     \node at (7.5,-3.6)  (q3232){};
     \node at (7.5,-4.2)  (q332){};
     \node at (7.5,-4.8)  (q44){}; 
     \node at (7.5,-5.8)  (q45){}; 
     \node at (7.5,-6.4)  (q46){}; 
     \node (end3) at (11,-3) {} edge [-] (q33);
     \node (end32) at (11,-3.6) {} edge [-] (q3232);
     \node (end32) at (11,-4.2) {} edge [-] (q332);
     \node (end4) at (11,-4.8) {} edge [-] (q44); 
     \node (end4) at (11,-5.8) {} edge [-] (q45); 
     \node (end4) at (11,-6.4) {} edge [-] (q46); 
     \node at (7.7,-5.8) (cr121) {$\times$};
     \node at (7.7,-6.4) (cr211) {$\times$};
     \node[ellipsis2] at (7.7,-5.8) (ep121) {};
     \node[ellipsis2] at (7.7,-6.4) (ep211) {};
     \draw[-] (ep121) -- (ep211);
     \node[ellipsis] (ee1) at (8.1,-5.2) {};
     \node[ellipsis] (ee2) at (8.1,-5.3) {};
     \node[ellipsis] (ee3) at (8.1,-5.4) {}; 
     \node at (8.5,-4.2) (cr121) {$\times$};
     \node at (8.5,-4.8) (cr211) {$\times$};
     \node[ellipsis2] at (8.5,-4.2) (ep121) {};
     \node[ellipsis2] at (8.5,-4.8) (ep211) {};
     \draw[-] (ep121) -- (ep211);
     \node at (9,-3.6) (cr121) {$\times$};
     \node at (9.,-4.2) (cr211) {$\times$};
     \node[ellipsis2] at (9,-3.6) (ep121) {};
     \node[ellipsis2] at (9,-4.2) (ep211) {};
     \draw[-] (ep121) -- (ep211);
     \node[operator3] (op21) at (10,-3.3) {$e^{-i S \Delta t}$} ;
     \node[ellipsis] (ee1) at (10.7,-5.2) {};
     \node[ellipsis] (ee2) at (10.7,-5.3) {};
     \node[ellipsis] (ee3) at (10.7,-5.4) {}; 
     \node at (10.7,-3.6) (hy3232) {$|$};
     \node at (10.7,-4.2) (hy332) {$|$};
     \node at (10.7,-4.8) (hy44) {$|$};
     \node at (10.7,-5.8) (hy44) {$|$};
     \node at (10.7,-6.4) (hy44) {$|$};
   \end{tikzpicture} } 
   \caption{Conventional method to realize $ e^{-i \rho t}$, where $\rho$ is a hermitian matrix. (a) is via decompositions from lie-trotter products and (b) is the density matrix exponentiation method using infinitesimal swaps. Wherein, $/$ denotes a register of multi qubits, and $|$ means tracing out the corresponding register.}   
\label{qPCA_original} 
\end{figure}
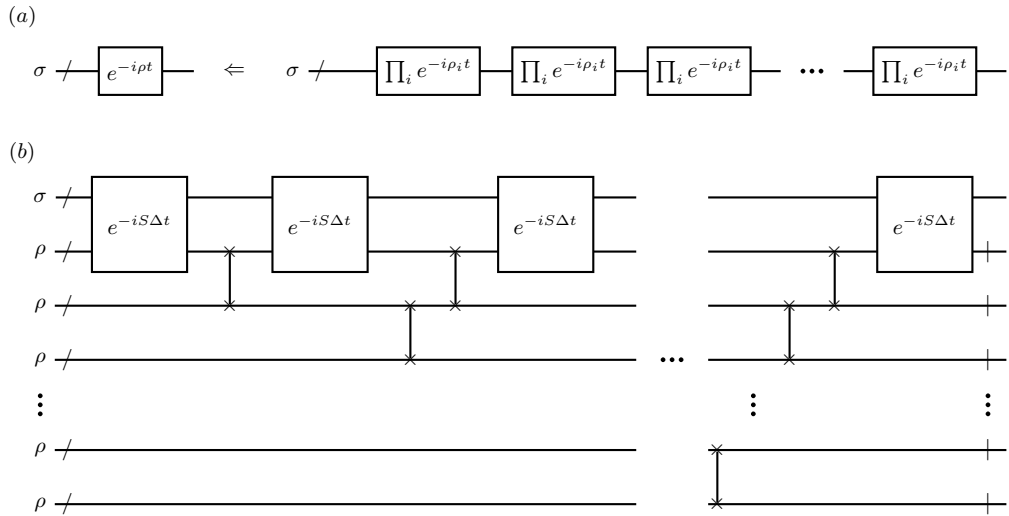

\emph{Lie-Trotter product and density matrix exponentiation} provide complete solutions for efficiently simulating matrix exponentiation where the matrices are hermitian. 
Traditionally, $e^{-i\rho t}$, where $\rho$ is either density matrix or Hamiltonian, can be simulated with Lie-trotter product formula.
\begin{eqnarray}\label{simulation_trotter}
   (\prod_i e^{-i\rho_i \Delta t})^n\rightarrow e^{-i\rho t},
\end{eqnarray}
where $\rho=\sum_i{\rho_i}$ and $\Delta t=t/n$. Fig.\ref{qPCA_original}(a) shows main process, where $n$ is the repeated times, chosen as $\mathcal{O}(t^2\epsilon^{-1}|\rho|^2)$ to guarantee an accuracy of $\epsilon$. 
Therefore, the depth of the circuit is at least $\mathcal{O}(t^2\epsilon^{-1})$ with a constant multiplied factor. 

Though repeated applications of simulated circuit is feasible in theory, a low-depth quantum circuit is preferred by current NISQ hardwares as their limited coherence time. Therefore, under the circumstance of a tolerant accuracy, the implementation appears unfriendly to current or near term devices.

Alternatively, density matrix exponentiation $e^{-i\rho t}$ can also be  simulated with a consumption of multiple copies of $\rho$ and infinitesimal swap operations.
Assuming that $\sigma$ is the density matrix of the system, the infinitesimal swap operation has such effect,
\begin{eqnarray}
   \mbox{tr}_p e^{-i S \Delta t} \rho\otimes \sigma e^{i S \Delta t}=\sigma-i\Delta t \left[\rho, \sigma\right] +\mathcal{O}(\Delta t^2), \label{erho}
\end{eqnarray}
where $\mbox{tr}_p$ is the partial trace over $\rho $ and $S$ is the swap operator. Shown in Fig.\ref{qPCA_original}(b), density matrix exponentiation with respect to $\sigma$ can be constructed by repeated applications of (\ref{erho}) with $n=t/\Delta t$ copies of $\rho$. Noted that
\begin{eqnarray}
   \underbrace{\mbox{tr}_{p}[e^{-i S \Delta t} \rho \otimes \mbox{tr}_{p}[e^{-i S \Delta t}\rho \otimes \cdots
      \mbox{tr}_{p}[}_{n} e^{-i S \Delta t} \rho \otimes \sigma e^{i S \Delta t}]\cdots e^{i S \Delta t} ] e^{i S \Delta t} ]=e^{-i \rho t} \rho e^{i \rho t} + \mathcal{O}(t^2n^{-1}),
\end{eqnarray}
where a guaranteed accuracy $\epsilon$ is determined by $n$, which decides both the depth and size of the circuit for simulating a $t=n\Delta t$-time evolution. Therefore, if the infinitesimal swap operation and related swap operations are defined as a layer of circuit, the depth is thus $\mathcal{O}(t^2\epsilon^{-1})$, which is also the size of circuit.

Though each layer only consists of infinitesimal swap operations, which is probably shallower than the previous one, the depth and the size of the circuit still dissatisfy the characteristics of nowadays NISQ hardwares, and thus hinder its near term application.

\begin{figure}[!h]
   \centering
   \begin{minipage}{0.5\textwidth}
      \resizebox{1\textwidth}{!}{%
    \begin{tikzpicture}[thick]
      \tikzstyle{operator2}=[draw,fill=white, text width=1.1cm, minimum height=4cm] 
    \node at (2,0.4) (q1) {$(a)$};
       \node at (2,0) (q1) {};
       \node at (2,-1) (q2) {};
       \node at (2,-2) (q3) {};
       \node at (2,-3.5) (q4) {};
       \node at (6.5,0) {} edge [-] (q1);
       \node at (6.5,-1){} edge [-] (q2);
       \node at (6.5,-2)  {} edge [-] (q3);
       \node at (6.5,-3.5) {} edge [-] (q4);
       \node at (8,0) (q5) {};
       \node at (8,-1) (q6) {};
       \node at (8,-2) (q7) {};
       \node at (8,-3.5) (q8) {}; 
       \node[meter] (end1) at (10.5,0) {} edge [-] (q5);
       \node[meter]  (end2) at (10.5,-1) {} edge [-] (q6);
       \node[meter]  (end3) at (10.5,-2) {} edge [-] (q7);
       \node[meter]  (end3) at (10.5,-3.5) {} edge [-] (q8);;
       \node[operator2] (op221) at (3.5,-1.75) {\small{$U_1(\theta_1)$}} ;  
       \node[operator2] (op222) at (5.5,-1.75) {\small{$U_2(\theta_2)$}} ;
      \node[ellipsis] (ee21) at (2.5,-2.5) {};
      \node[ellipsis] (ee22) at (2.5,-2.75) {};
      \node[ellipsis] (ee23) at (2.5,-3) {}; 
       \node[ellipsis] (ee21) at (7,-0) {};
       \node[ellipsis] (ee22) at (7.25,-0) {};
       \node[ellipsis] (ee23) at (7.5,-0) {};    
       \node[ellipsis] (ee21) at (7,-1) {};
       \node[ellipsis] (ee22) at (7.25,-1) {};
       \node[ellipsis] (ee23) at (7.5,-1) {};    
       \node[ellipsis] (ee21) at (7,-2) {};
       \node[ellipsis] (ee22) at (7.25,-2) {};
       \node[ellipsis] (ee23) at (7.5,-2) {};    
       \node[ellipsis] (ee21) at (7,-3.5) {};
       \node[ellipsis] (ee22) at (7.25,-3.5) {};
       \node[ellipsis] (ee23) at (7.5,-3.5) {};    
       \node[ellipsis] (ee21) at (7.25,-2.5) {};
       \node[ellipsis] (ee22) at (7.25,-2.75) {};
       \node[ellipsis] (ee23) at (7.25,-3) {}; 
       \node[ellipsis] (ee21) at (10,-2.5) {};
       \node[ellipsis] (ee22) at (10,-2.75) {};
       \node[ellipsis] (ee23) at (10,-3) {}; 
       \node[operator2] (op226) at (9,-1.75) {\small{$U_m(\theta_m)$}} ;  
      \end{tikzpicture}  } 
\end{minipage}
\quad
   \begin{minipage}{0.5\textwidth}
   \resizebox{1\textwidth}{!}{%
   \begin{tikzpicture}[thick]
   \ctikzset{scale=1.5}
   \tikzstyle{every node}=[font=\normalsize,scale=1]
     \tikzstyle{operator} = [draw,fill=white,minimum size=1em] 
     \tikzstyle{operator2} = [draw,fill=white,minimum size=2em] 
     \tikzstyle{operator3} = [draw,fill=white,minimum size=4em] 
     \tikzstyle{operator4} = [draw,shape=rectangle, fill=white, minimum width=1em, minimum height=10em] 
     \tikzstyle{phase} = [fill,shape=circle,minimum size=3pt,inner sep=0pt]
     \tikzstyle{surround} = [fill=blue!10,thick,draw=black,rounded corners=2mm]
     \tikzstyle{ellipsis} = [fill,shape=circle,minimum size=2pt,inner sep=0pt]
     \tikzstyle{ellipsis2} = [fill,shape=circle,minimum size=0.5pt,inner sep=0pt]
     \node at (2.2,-0.5) (q2) {$(b)$};
     \node at (2.2,-1) (q2) {$S$};
     \node (end3) at (9,-1) {} edge [-] (q2);
     \node at (2.5,-1) (hy111) {$/$};
     \node at (2.2,-2) (q3) {$A$};
     \node at (2.3,-2.5) (rho2) {$\Omega_0$};
     \node (end3) at (9,-2) {} edge [-] (q3);
     \node[meter] (op22) at (9,-1) {} ;
     \node[meter] (op22) at (9,-2) {} ;
     \node at (2.5,-2) (hy111) {$/$};
     \node[operator2] (op22) at (3,-2) {$H$} ;
     \node[operator2] (op22) at (3,-1) {$H$} ;
     \node[operator2] (op22) at (4,-2) {$H$} ;
     \node[phase] (ph21) at (3.5,-1) {};
     \node[phase] (ph22) at (3.5,-2) {} edge [-] (ph21);
     \draw [dashed] (4.5,-0.7) -- (4.5,-2.3);
     \node at (4.55,-2.5) (rho1) {$\Omega_1$};
     \node[operator2] (op22) at (5.5,-1) {\small{$U_{PQC}(\bm{\theta})$}} ;
     \draw [dashed] (6.5,-0.7) -- (6.5,-2.3);
     \node at (6.55,-2.5) (rho2) {$\Omega_2$};
     \node[operator2] (op22) at (7,-2) {$H$} ;
     \node[operator2] (op22) at (8,-1) {$H$} ;
     \node[operator2] (op22) at (8,-2) {$H$} ;
     \node[phase] (ph11) at (7.5,-1) {};
     \node[phase] (ph12) at (7.5,-2) {} edge [-] (ph11);
     \draw [dashed] (8.65,-0.7) -- (8.65,-2.3);
     \node at (8.7,-2.5) (rho2) {$\Omega_3$};
     \begin{pgfonlayer}{background} 
      \node[rectangle, fill=blue!10, rounded corners=3mm, minimum height=3cm, opacity=.8, text width=2.25cm, align=left, outer sep=5mm] (main) at (3.5,-1.5) {};
      \node[rectangle, fill=pink!35, rounded corners=3mm, minimum height=3cm, opacity=.9, text width=2.cm, align=left, outer sep=5mm] (main) at (5.5,-1.5) {};
       \node[rectangle, fill=green!20, rounded corners=3mm, minimum height=3cm, opacity=.9, text width=2.25cm, align=left, outer sep=5mm] (main) at (7.5,-1.5) {};
       \node at (3.5,-2.5) (qqq) {$U_E$};
       \node at (5.5,-2.5) (qqq) {$U_{PQC}\otimes I$};
       \node at (7.5,-2.5) (qqq) {$U_D$};
      \end{pgfonlayer} 
   \end{tikzpicture}}      
   \end{minipage}
   \caption{Parameterized quantum circuit and ancilla-assist parameterized quantum circuit. (a) gives an example of a $m$-layer parameterized circuit, which is employed in (b) as $U_{PQC}(\bm{\theta})$. $U_i$ are parametrized quantum gates with tuneable $\theta_i$. 
   (b) is the ancilla-assist parameterized quantum circuit, which has three parts: encoding circuit(blue) $U_E$, parameterized circuit(pink) $U_{PQC}\otimes I$ and decoding circuit(green) $U_D$. $H$ are a bunch of Hadamard gates, as well as control-z gates.}   
\label{APQC_demo} 
\end{figure}
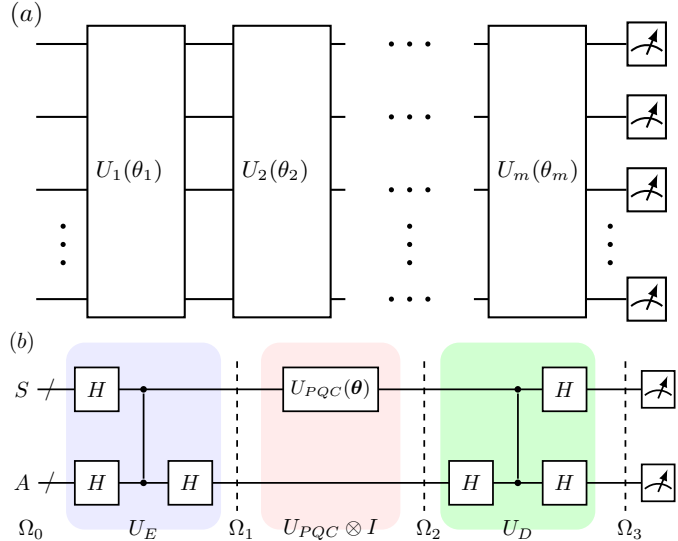

\emph{Parameterized quantum circuit} is an important part for the NISQ algorithms, providing a concrete way to implement algorithms in the NISQ era, such as quantum approximation  optimization algorithm  and variant quantum eigen-solver\cite{farhi2014quantum,peruzzo2014variational}.
A typical $m$-layer PQC is schematically presented in Fig.\ref{APQC_demo}(a). It can be mathematically expressed as $U(\bm{\theta})=\prod_{i=1}^N U_i(\theta_i)$, where $\theta_i$ is tuneable parameters for $U_i$, the $i$-th layer unitary operation. 
As the NISQ devices are with characteristics such as, limited qubits, limited coherence time, and accurate one or two qubits quantum operation, a limited depth PQC seems perfect candidate for comprising a NISQ algorithm. Additionally, although only limited depths circuits and basic quantum gates are utilized in PQCs, the expressivity are investigated and PQCs are still good at presenting non-trivial quantum states\cite{nakaji2020expressibility,sim2019expressibility}. 

Therefore, for important roles played towards the realization of the NISQ algorithms, in our framework, PQCs are employed to approach the simulations of matrix exponentiation, which pushes our framework to near future devices. Typically, in PQC, $U_i$ can be composed of entanglement gates and single qubit gates, whose structure generally depends on the tasks at hand. 
Problem-inspired circuit is one type, employing information about the problem, while some problem-agnostic circuits works even when no relevant information is previously known. The widely employed structures include quantum alternating operator ansatz, variational Hamiltonian ansatz and unitary coupled clustered ansatz.

From the point of hardware efficiency, in this work, one kind of $N$-layer circuit is designed as a trial for the following $N$-qubit tasks, where all two body interactions are included and gates in one layer are commuted with each others. which aims at employing the two-body interactions of the devices with a limited depth circuit. The method generating this structure is specified in supplementary\cite{supp}. As for finding appropriate parameter configuration for the PQC, two strategies are provided, using a newly proposed circuit paradigm introduced as follows. Remarkably, the proposed structure of circuit can be changed with the task and two strategies are applicable to a general PQC.

\emph{Ancilla-assisted Parametrized Quantum Circuit(APQC)} is a generalized PQC with ancillary systems, which is schematically depicted in Fig.\ref{APQC_demo}(b). 
Besides the working system which is denoted as $S$,  the ancillary register $A$ is appended with a same size as $S$.
For initialization, they are jointly prepared on $\Omega_0=\ket{\psi_0}\bra{\psi_0}$, where
\begin{eqnarray}
   \ket{\psi_0}=\ket{0}_S\ket{0}_A.
\end{eqnarray}
The entire process of the circuit includes three parts: encoding circuit($U_E$, colored blue), PQC part($U_{PQC}\otimes I$, colored pink), and decoding circuit($U_D$, colored green).

To make use of Chio-Jamiolkowski isomorphism, encoding part $U_E$ is supposed to evolve the system into $\Omega_1=\ket{\psi_1}\bra{\psi_1}$, where $|\psi_1\rangle=\sum_{i}\ket{i}_S\otimes\ket{i}_A$ are pairs of bell states. This step can be realized using a bunch of control-z gates, $C_z$ and Hadamard gates, $H_A$, which is expressed as
\begin{eqnarray} 
   \Omega_1&=&U_E (\Omega_0), \nonumber \\
   U_E&=&I_S \otimes H_A \cdot Cz \cdot H_S \otimes H_A.
\end{eqnarray}

Then, the parametrized circuit $U_{PQC}$ is applied on register $S$, driving the system into $\Omega_2$, which is also called the choi matrix of $U_{PQC}$, 
\begin{eqnarray} 
   \Omega_2=\ket{\psi_2}\bra{\psi_2} = U_{PQC}\otimes I (\ket{\psi_1}\bra{\psi_1}) = \sum_{k,l} U_{PQC}(\ket{k}\bra{l})_S\otimes \ket{k}\bra{l}_A.\label{omega_2}
\end{eqnarray}
As arbitrary state $\rho$ with the application of $U_{PQC}$ has such state-channel duality,
\begin{eqnarray}
   U_{PQC}(\rho)=Tr_A[\Omega_2\cdot I \otimes \rho^*],
   \label{processbasis}
\end{eqnarray}
$\Omega_2$ is thus sufficient to determining $U_{PQC}$, which implies the feasibility to depict a quantum channel via state characterization. Furthermore, Choi-Jamiolkowski isomorphism is exactly the correspondences in Eqs. (\ref{omega_2}) and (\ref{processbasis}).

Finally, $\Omega_{3}$ is the result generated by steering $\Omega_{2}$ through the decoding circuit $U_D$, where 
\begin{eqnarray}
   U_D=H_S \otimes H_A \cdot Cz \cdot I_S \otimes H_A.
\end{eqnarray}
It is the inverse of the encoding circuit, which is employed as a preliminary treatment for the measurements and can be modified with variant problems. In our task, the decoding circuit disentangles the system which is essential for the second strategy.

\emph{The first strategy} provides a method, generating an appropriate PQC, which simulates a $t$-time evolution, $e^{-it\rho}$, where $\rho$ is denoted as a hermitian matrix.
Specifically, if $U_i(\theta_i)$ of Fig.\ref{APQC_demo}(a) are substituted with $e^{-it \alpha_i \rho_i}$, where $\rho_i$ are tensor products of Pauli operators, the problem is thus stated as approaching $e^{-it\rho}$ with a series of evolution segments,
\begin{eqnarray}
   e^{-it \alpha_m\rho_m}\cdots e^{-it \alpha_1 \rho_1},
\end{eqnarray}   
which compose a parameterized quantum circuit. For the choices of $\rho_i$, i.e., determining the structure of PQC, it is an empirical task and depends on the task at hand. In this task, the $N$-layer circuit is employed, which is depicted as a trial. Furthermore, with the optimal control theory, configurations of $\alpha_i$ can be optimized and can be iteratively found. 
From the point of view of state, for a specified time $t$, $\rho$ evolves an arbitrary system, which is initialized at $\sigma$, into
\begin{eqnarray}
   \sigma_0(t) =e^{-it\rho}\sigma e^{it\rho}.
\end{eqnarray}
Simultaneously, with the application of PQC, the system is steered to $\sigma(t)$, where
\begin{eqnarray}
   \sigma(t) =\prod_{i=1}^m e^{-it \alpha_i \rho_i}\sigma \prod_{i=1}^m e^{it \alpha_i \rho_i}.
\end{eqnarray}
The objective function $f$ is thus defined as the overlap which is measured by the standard inner product 
\begin{eqnarray}
   f(\bm{\alpha})&=&tr(\sigma_0(t)\sigma(t)) =\langle e^{-it\rho}\sigma e^{it\rho}e^{-it \alpha_m\rho_m}\cdots e^{-it \alpha_1 \rho_1} \sigma e^{it \alpha_1\rho_1}\cdots e^{it \alpha_m \rho_m}\rangle\nonumber\\
   &=&\underbrace{\langle e^{it \alpha_{k+1} \rho_{k+1}}\cdots e^{it \alpha_m \rho_m}e^{-it\rho}\sigma e^{it\rho}e^{-it \alpha_m\rho_m}\cdots e^{-it \alpha_{k+1} \rho_{k+1}}}_{\mu_k}| \underbrace{e^{-it \alpha_k \rho_k} \cdots e^{-it \alpha_1 \rho_1} \sigma e^{it \alpha_1\rho_1}\cdots e^{it \alpha_k \rho_k}}_{\nu_k}\rangle.
   \label{obj_algo1}
\end{eqnarray}
$f(\bm{\alpha})\leq 1$ and the maximum case corresponds to a state to state case, where the $m$-layer PQC realizes a target state evolved by $\rho$ for a specified time $t$.

As for parameter optimizations, a conventional method to optimizing the objective function is based on gradient, where the partial derivative with respect to the parameter $\alpha_k$ are 
\begin{eqnarray}
   \frac{\partial f}{\partial{\alpha_k}}=-it  \langle \mu_k| [\rho_k,\nu_k]\rangle.
   \label{graident1}
 \end{eqnarray}
Therefore, by updating $\alpha_i$ with $\alpha_i'$ with a learning rate $\eta$, where
\begin{eqnarray}
   \alpha_k'\rightarrow \alpha_k+\eta \frac{\partial{f}}{\partial{\alpha_k}},\label{alpha_updated}
\end{eqnarray}
the objective function would increase along the direction of anti-gradient until converging into the local maximum. If $f\approx 1$ is iteratively achieved, this PQC is said to be able to drive the system $\sigma$ to $\sigma_0(t)$, which is the application of $e^{-it\rho}$. 

Noticeably, this heuristic method seeks a state to state method. For generating a quantum operation approaching the evolution of $\rho$, ancilla-assistant parameterized quantum circuit(APQC) is introduced, transforming the state to state method to gate based optimization method.
To specify the $U_{PQC}$ with $\prod_{i=1}^N e^{-it \alpha_i \rho_i}$,  the notations in Eq.(\ref{obj_algo1}) should be updated as, 
\begin{eqnarray}\label{core_algorithm1}
   \sigma&=& U_E \Omega_0 U_E^{\dagger},\nonumber \\
   \mu_k&=&\prod_{i=k+1}^m e^{it \alpha_{i} \rho_{i}}\otimes I\cdot e^{-it\rho}\otimes I \left(\sigma\right)\nonumber \\
   \nu_k&=&\prod_{i=k}^1 e^{it \alpha_{i} \rho_{i}}\otimes I (\sigma)
\end{eqnarray}
where $\alpha_i$ are to be optimized by (\ref{alpha_updated}). 
Obviously, this strategy has one vital problem. It lies at calculating the gradient, which is in general inefficient as obtaining $\nu_k$ and $\mu_k$ is time consuming, and cannot be dealt with existed methods for most cases\cite{mitarai2018quantum}. \emph{the second strategy} is thus reported.

\emph{The second strategy} is outlined as follows, (i) implementing the simulation of $e^{i\rho \Delta t}$, where $\Delta t$ is a sufficiently short period of time; (ii) implementing $e^{i\rho t}$ using APQC based on $e^{i\rho \Delta t}$ step by step, where $\mathcal{O}(\log(t/\Delta t))$ steps are required. This idea is extended from Eq.(\ref{simulation_trotter}), which guarantees a trotter error with $\epsilon_t\sim\mathcal{O}(t\Delta t)$.
  
For simulating $e^{-i\rho \Delta t}$, i.e.,
\begin{eqnarray}
\prod_i e^{-i\Delta t\beta_i\rho_i} \Rightarrow  e^{-i\Delta t\rho},
\end{eqnarray}
$\rho_i$ and $\bm{\beta}$ should be determined. In the point of view of an arbitrary state, $\sigma_0$, which is employed to initialize the system, an objective function can thus be constructed as
\begin{eqnarray}
   f(\bm{\beta})=\langle e^{-i\Delta t\rho}\sigma e^{i\Delta t\rho}\prod_i e^{-i\Delta t\beta_i\rho_i} \sigma \prod_i e^{i\Delta t\beta_i\rho_i}\rangle.
\end{eqnarray}
Noticeably, $f(\bm{\beta})\leq 1$ as all operations are supposed as unitary. Expanding $e^{-i\Delta t\rho}$ and $\prod_{i=1} e^{-i\Delta t \beta_i \rho_i}$ with Taylor decompositions on $\Delta t$, the objective function has two parts, $1$ and a power series representation on $\Delta t$, which can be employed as a metric to measure the discrepancy between the target state and the result of PQC. Though it will be more accurate as more terms are involved, for a sufficiently small $\Delta t$, this power series representation can be approximated with only the lowest order of $\Delta t$.
Coincidentally, 
\begin{eqnarray}
   \sum \beta_i\rho_i=\rho.
   \label{Trotter-decom1}
\end{eqnarray}
is the condition eliminating the lowest order term, which is exactly the result from Lie-Trotter decomposition, providing a method for simulating a sufficiently small period time evolution. Related details for derivation can be found in Appendix\cite{supp}. 
Therefore, assumed that $\Delta t$ is small enough, $e^{i\rho \Delta t}$ can be simulated efficiently, using a parameterized circuit, where structure and parameters are given by Eq.(\ref{Trotter-decom1}).

\begin{figure}[!h]
   \centerline{
   \begin{tikzpicture}[thick]
   \ctikzset{scale=1.4}
   \tikzstyle{every node}=[font=\normalsize,scale=1]
     \tikzstyle{operator} = [draw,fill=white,minimum size=1em] 
     \tikzstyle{operator2} = [draw,fill=white,minimum size=2em] 
     \tikzstyle{operator3} = [draw,fill=white,minimum size=4em] 
     \tikzstyle{operator4} = [draw,shape=rectangle, fill=white, minimum width=1em, minimum height=10em] 
     \tikzstyle{phase} = [fill,shape=circle,minimum size=3pt,inner sep=0pt]
     \tikzstyle{surround} = [fill=blue!10,thick,draw=black,rounded corners=2mm]
     \tikzstyle{ellipsis} = [fill,shape=circle,minimum size=2pt,inner sep=0pt]
     \tikzstyle{ellipsis2} = [fill,shape=circle,minimum size=0.5pt,inner sep=0pt]
     \node at (2.2,-2) (q2) {$S$};
     \node (end3) at (6.8,-2) {} edge [-] (q2);
     \node at (7.2,-2) (q2) {};
     \node (end3) at (10.3,-2) {} edge [-] (q2);
     \node at (2.5,-2) (hy111) {$/$};
     \node at (2.2,-3) (q3) {$A$};
     \node at (2.3,-3.5) (rho2) {$\Omega_0$};
     \node (end3) at (10.3,-3) {} edge [-] (q3);
     \node at (2.5,-3) (hy111) {$/$};
     \node[operator2] (op22) at (3,-3) {$H$} ;
     \node[operator2] (op22) at (3,-2) {$H$} ;
     \node[operator2] (op22) at (4,-3) {$H$} ;
     \node[phase] (ph21) at (3.5,-2) {};
     \node[phase] (ph22) at (3.5,-3) {} edge [-] (ph21);
     \draw [dashed] (4.5,-1.7) -- (4.5,-3.3);
     \node at (4.55,-3.5) (rho1) {$\Omega_1$};
     \node[operator2] (op22) at (5.25,-2) {\small{$U_{i+1}(\bm{\beta})$}} ;
     \node[operator2] (op22) at (6.5,-2) {$U_{i}^{-1}$} ;
     \node[ellipsis] (ee21) at (6.9,-2) {};
     \node[ellipsis] (ee22) at (7,-2) {};
     \node[ellipsis] (ee23) at (7.1,-2) {}; 
     \node[operator2] (op22) at (7.5,-2) {$U_{i}^{-1}$} ;
     \draw [dashed] (8,-1.7) -- (8,-3.3);
     \node at (8,-3.5) (rho2) {$\Omega_2$};
     \node[operator2] (op22) at (8.5,-3) {$H$} ;
     \node[operator2] (op22) at (9.5,-2) {$H$} ;
     \node[operator2] (op22) at (9.5,-3) {$H$} ;
     \node[phase] (ph11) at (9,-2) {};
     \node[phase] (ph12) at (9,-3) {} edge [-] (ph11);
     \node at (10.2,-3.5) (rho2) {$\Omega_3$};
     \begin{pgfonlayer}{background} 
      \node[rectangle, fill=blue!10, rounded corners=1mm, minimum height=3cm, opacity=.8, text width=2.25cm, align=left, outer sep=5mm] (main) at (3.5,-2.5) {};
      \node[rectangle, fill=pink!35, rounded corners=3mm, minimum height=3cm, opacity=.9, text width=4.25cm, align=left, outer sep=5mm] (main) at (6.25,-2.5) {};
       \node[rectangle, fill=green!20, rounded corners=3mm, minimum height=3cm, opacity=.9, text width=2.25cm, align=left, outer sep=5mm] (main) at (9,-2.5) {};
       \node at (3.5,-3.5) (qqq) {$U_E$};
       \node at (6.25,-3.5) (qqq) {$\xi\otimes I$};
       \node at (9,-3.5) (qqq) {$U_D$};
      \end{pgfonlayer} 
   \end{tikzpicture} } 
   \caption{Ancilla-assist parameterized quantum circuit to implement \emph{algorithm 2}. $U_{i+1}(\bm{\beta})$ is the circuit to be optimized with known parameterized circuit $U_i$. Notations are the same as Fig.\ref{APQC_demo}(a).}   
\label{repeating algorithm} 
\end{figure}
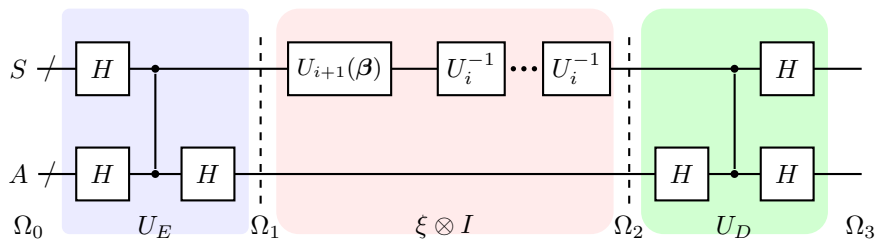

For simulating $e^{i\rho t}$, APQC should be repeatedly conducted for $\mathcal{O}(\log(t/\Delta t))$ times.
The strategy for the $i$-th conduction is depicted in Fig.\ref{repeating algorithm}, where $U_{PQC}$ consists of $U_{i+1}(\bm{\beta})$ and $U_{i}^{-1}$ followed for $n_c$ times. Remarkably, $n_c$ represents the compression efficiency and $U_{i}$ should be well parameterized.
Therefore, the remaining task for \emph{the second strategy} is to generate $U_{i+1}(\bm{\beta})$ with an explicitly prior known parameterized circuit $U_{i}$, where $U_{i+1}=U_{i+1}^{n_c}$ and $\bm{\beta}$ are the parameters to be optimized.

Firstly, an objective function of $\bm{\beta}$, is defined as,
\begin{eqnarray}\label{obj_APQC}
   f=tr(\Omega_1\Omega_{2})=tr(\Omega_0\Omega_{3}),
\end{eqnarray}
which reaches the maximum at $U_{i+1}(\bm{\beta})=U_i^{n_c}$. As $\Omega_1$ are pairs of bell states, this objective function cannot be measured with the local bases directly and efficiently. A decoding circuit $U_D$ disentangles the target component into $\Omega_0$. In this situation, only with the local measurement on $\ket{\psi_0}$, the objective function can be estimated. 

As for optimizing parameters in above circuit, it is allocated to the classical computer which is similar as other variant quantum algorithms. The common methods include mainly two classes, gradient-based approaches and gradient free approaches, which have been investigated in plenty of algorithm engineering tasks\cite{lavrijsen2020classical}.

In this way, the parameters in $U_{i+1}$ which is well parameterized and obtained in the $i$-th conduction of APQC, could be applied in the $i+1$-th step, where $U_{i+2}=U_{i+1}^{n_c}$ is realized. Accordingly, by $\log(t/\delta)$ steps, $e^{it\rho}$ can be implemented based on $e^{i\Delta t\rho}$. Compared to repeatedly applications of $U_{i}$ which will cause the circuit depth affordable, this strategy trades the circuit depth with applications of APQC, which increases the training time while compresses the repeated circuits. for many repeated applications in quantum phase estimation and its generalization, the circuit could be modified with APQC into a more Experimental friendly form.

\section{Analysis}

In this section, the consumptions for framework are studied as well as an analysis on errors and expressibility for generated parameterized quantum circuit. 

For the complexity of our framework, it is divided into two part: simulation complexity and strategy complexity, which generated the PQCs for simulation.
Simulation complexity are the same for both strategies, as a limited depth circuit is purposeful generated with at most two-qubit parametrized quantum operations.

\begin{minipage}{0.5\textwidth}
   \begin{algorithm}[H]
      \begin{flushleft}
      \caption{}\label{tab_algorithm1}
       \hspace{\algorithmicindent} \textbf{Input:} parameterized circuit $\prod_{i=1} e^{-it \alpha_i \rho_i}$, which is used in APQC.
       $\bm{\alpha}$ is to be optimized, $\epsilon_o$ is an optimized threshold, and $\delta_1$ is the tolerance for improving.\\
       \hspace{\algorithmicindent} \textbf{Output:} $\bm{\alpha}$, parameter configuration, which is optimized to approach $e^{-it\rho}$ via $\prod_{i=1} e^{-it \alpha_i \rho_i}$.
      \end{flushleft}
       \begin{spacing}{1.3}
         \begin{algorithmic}[1]
         \State Evaluate the objective function of APQC with existed $\bm{\alpha}$, denoted as $f$. If $f\leq 1-\epsilon_o$, go to 2, otherwise, the algorithm terminates and $\bm{\alpha}$ return. 
         \For{$k=1,...,m$}
         \State Based on Eq.(\ref{core_algorithm1}), $\nu_k$ is calculated with $\sigma$ and $\mu_k$ is calculated with $(e^{-it\rho}\otimes I)\sigma(e^{-it\rho}\otimes I)^{\dagger}$.
         \State Evaluate $\partial f/\partial \alpha_k$ by Eq.(\ref{graident1}).
         \EndFor
         \State Update $m$-element $\bm{\alpha}$ according to Eq.(\ref{alpha_updated})
         \State Evaluate the objective function of APQC with the new $\bm{\alpha}$
         \If{$f\geq 1-\epsilon_o$} 
         \State the algorithm terminates and $\bm{\alpha}$ return;
         \ElsIf{$\Delta f \leq \delta_1$}
         \State $\bm{\alpha}$ are re-initialized and go to 1.
         \Else
         \State go to 2.
         \EndIf
         \\
         \\
         \end{algorithmic}
        \end{spacing}
   \end{algorithm}
   \end{minipage}
   \hfill
   \begin{minipage}{0.5\textwidth}
   \begin{algorithm}[H]
      \begin{flushleft}
         \caption{}\label{tab_algorithm2}
       \hspace{\algorithmicindent} \textbf{Input:} 
       parameterized circuit $\prod_{i=1} e^{-i \beta_i \rho_i}$ with known $U_1$. $\bm{\beta}$ is to be optimized, $\epsilon_o$ is the optimized threshold, $\delta_2$ is the tolerance for improving, and $n_c$ is the compressing factor. \\
       \hspace{\algorithmicindent} \textbf{Output:}  $\bm{\beta}$, parameter configuration, which is optimized to approach $e^{-it\rho}$ via $\prod_{i=1} e^{-it \beta_i \rho_i}$
      \end{flushleft}
       \begin{spacing}{1.3}
         \begin{algorithmic}[1]
         \For{$i=1,...,\log(t/\Delta t)$}
         \If{$i=1$}
         \State  $U_1$ and its inversion are generated by the input.
         \Else
         \State  $U_i$ are generated as the output of the last iteration.
         \EndIf
         \State Evaluate the objective function of APQC with existed $\bm{\beta}$ and $n_c$, which is denoted as $f$. If $f\leq 1-\epsilon_o$, go to 8, otherwise, $\bm{\beta}$ for $U_{i+1}$ return and go to 1. 
         \State $\bm{\beta}$ for $U_{i+1}$ are optimized as variant quantum algorithms, $\bm{\beta}$ are updated and $f$ is evaluated.
         \If{$f\geq 1-\epsilon_o$} 
         \State $\bm{\beta}$ return and go to 1;
         \ElsIf{$\Delta f \leq \delta_2$}
         \State $\bm{\beta}$ are re-initialized and go to 7.
         \Else
         \State go to 8.
         \EndIf
         \EndFor
         \end{algorithmic}
        \end{spacing}
   \end{algorithm}
   \end{minipage}
   \\
   \\

As for the strategy complexity, we list their procedures in the Strategy.\ref{tab_algorithm1} and \ref{tab_algorithm2}, and corresponding complexities are analyzed as follows, where $r$ is the total iterations numbers for optimizations. 
As APQC is employed, $2n$ qubits are required for a $n$-qubit simulation problem, which is the memory complexity. 
As for the gate complexity, the complexities introduced by APQC is as follows. $\mathcal{O}(n)$ Hadamard gates and $\mathcal{O}(n)$ control-Z gates are required for the encoding and decoding circuit. In addition, a depth of $\mathcal{O}(m)$ parameterized quantum circuit which consists of at most $\mathcal{O}(m\times n)$ local quantum gates and $\mathcal{O}(m\times n)$ two-qubit quantum gates. In this paper, a kind of PQC with all two body connections is employed as a prototype, which leads $m=\mathcal{O}(n)$. 

Additionally, for the procedure of training, \emph{the first strategy} mainly relies on classical simulation, which assists the optimization to generate parameterized quantum circuits.
During the strategy, approximating the gradient with Eq.(\ref{graident1}) requires the most expenditure of computation resource.
For each iteration, calculating $\mu_k$ and $\nu_k$ lies at the heart, which requires the implementation of matrix multiplications on a classical computer. Though $e^{-it\rho_i}$ can be efficiently simulated  individually for the locality of $\rho_i$, simulating $e^{-it\rho}$ is in general hard for most cases and there is no universal efficient algorithms. To simulate a $n$-qubit problem, $2^n \times 2^n$ matrix should be generated and stored, with $\mathcal{O}(4^n)$ operations for evaluating the objective function and obtaining the gradient. 
Thus, the entire operation cost is $\mathcal{O}(4^n)\times r$ for a total $r$ iterations. As for using APQC which doubles the space, the complexity for each iteration would be squared.

Therefore, \emph{the first strategy} is in general inefficient. However, it would be applicable to some circumstances where $\nu_k$ and $\mu_k$ can be efficiently simulated. In fact, some investigations have employed tensor network to work out this simulation problem and shed the light on the middle-scale quantum systems\cite{pan2021simulating}. 
Furthermore, this heuristic method can be generalized to realize an arbitrary quantum gate using PQC within a bounded error, by replacing $e^{-it\rho}$ with an target unitary operation.

For \emph{the second strategy}, updating all $m\times n$ tuneable parameters requires $(2\times m\times n)$ times repeating the APQC and measurements. Therefore, $\mathcal{O}(m\times n)\times r$ are time complexity for $r$ iterations.
Additionally, with the assumptions that parameters in APQC can be well stored, repeatedly applications of APQC can thus be realized with the accurate local operations for low-depth current devices, which is within the capability of near term quantum devices. Therefore, this repeating leads an additional multiplied factor, $\log{(t/\Delta t)}$, for the time complexity, which is at last $\mathcal{O}(m\times n\times r\times \log{(t/\Delta t)})$.

For analyzing errors of generated PQCs form both strategies, $\epsilon_o$ is assumed as the optimized threshold which is supposed to terminate the training process, $\epsilon_t$ is assumed as the deviations coming from lie-product decompositions. 
If \emph{the first strategy} is completed, the obtained PQC has an accurate of $1-\epsilon_o$ as training process permits the error no more than $\epsilon_o$.
However, for \emph{the second strategy} which has a less than $\epsilon_o$ deviation at each step implementing $U_{i+1}$ by $U_i$, the error accumulates according to a chain rule. It ends up with an error of $\mathcal{O}(n_c^s\epsilon_o)$, where the higher order terms are ignored. As $s$ is total steps required which is $\mathcal{O}(\log(t/\Delta t))$ for $e^{-i\rho t}$ from $e^{-i\rho \Delta t}$, the deviation is thus $\mathcal{O}((t/\Delta t)\epsilon_o)$. Additionally, an error coming from Eq.(\ref{simulation_trotter}) is also considered as $\epsilon_t$, which is of $\mathcal{O}(t\Delta t)$. Accordingly, the total error for \emph{the second strategy} is $\mathcal{O}((t^2/\epsilon_t)\epsilon_o)+\mathcal{O}(\epsilon_t)$.
   
   \begin{figure}[htb]
      \begin{center}
      \includegraphics[width= 0.6\columnwidth]{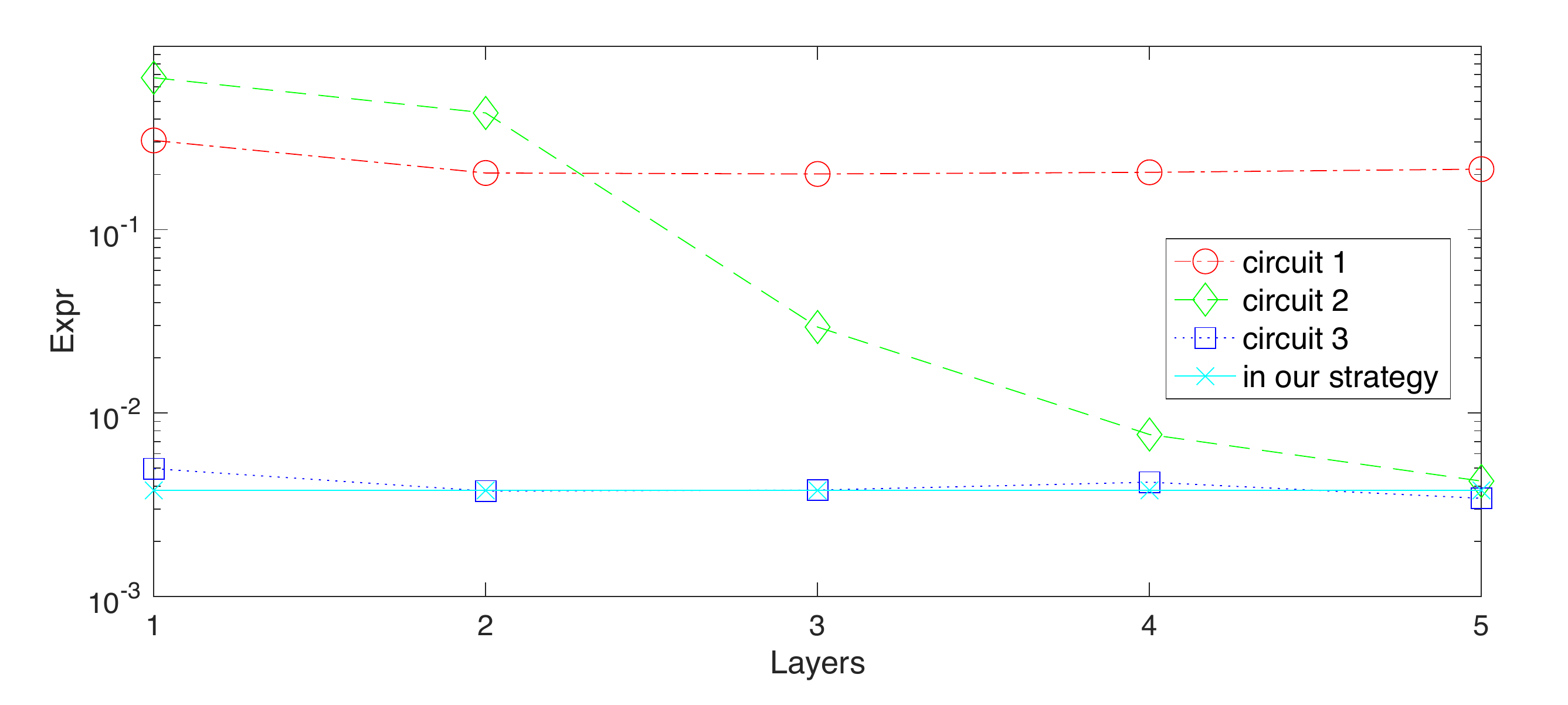}
      \end{center}
      \setlength{\abovecaptionskip}{-0.00cm}
      \caption{Expressibility values computed for circuits specified in Appendix and circuit used in our strategies. Circuit 1(red circles), 2(gree diamonds) and 3(blue squares) are repeatedly applied with up to 5, while the circuit employed in our strategy(cyan cross) keeps unchanged.}\label{expr}
   \end{figure}
  
   In our strategies, a structure of parameterized circuit with all two body interactions is employed as a prototype, whose details are introduced in Appendix. In fact, it is problem-dependent and can be  replaced with smarter ansatz with tasks at hand. Moreover, our strategies can achieve parameter optimizations for more kinds of circuit.    
   Nevertheless, the analysis on expressive power of such structure is provided.
   Expressibility is proposed recently as a distance, which measures the output states distribution of PQC and the Haar\cite{sim2019expressibility}. This could be used as a metric for expressive power of certain circuit, where a highly expressible circuit would produce a small expressibility value.
   To explicitly present this value,  Kullback-Leibler (KL) divergence(or relative entropy) is employed to estimate this distance, which is denoted as $Expr$. 
   In this part, besides numerically calculating $Expr$ of circuit employed in our strategy, three other types of parametrized circuits are also studied as comparisons, which are specified in Appendix\cite{sim2019expressibility}. 
   If one circuit is regarded as a unit layer, three compared circuits are repeated with up to 5 times to calculate $Expr$. While the circuit used in our strategies keeps unchanged.  
   Fig.\ref{expr} shows the results of Expressibility values (or KL divergences), where circle, diamond and square represents three comparison circuits and cyan cross labels the circuit used in our strategy. 
   In general, repeating a circuit layer would increase the expressive power.As there is no entanglement gate, this argument does not hold for circuit 1.
   From the result, we can see this trend, as well as their convergences of expressibility. 
   Therefore, with respect to the metric of expressibility, the PQC used in our strategy have a similar performance as multi-layer circuits 2 and 3.
   For more information on expressibility and its simulation, they can be found in Appendix or related work\cite{supp,sim2019expressibility}.

\section{Application and Numerical Experiments}

To mimic the dynamics of a quantum system with quantum computer, various simulation algorithms have been proposed. However, most of them employ a deep quantum circuits which is beyond the capability of near-term NISQ hardware.
As the natural application, our strategies can be employed in the field of Hamiltonian simulation, studying the dynamics using the PQCs.

Quantum systems with short range interactions are generally existed in physical world. In this specific situation, we consider a $N$-qubit quantum system with at most two-body interactions. 
\begin{eqnarray}
   H=\sum_{i=1}^N H_1(i)+\sum_{i>j,i,j=1}^N H_2(i,j),
\end{eqnarray}
where $H_1$ and $H_2$ are tensor products of Pauli matrices for one and two body interactions. 
The target is to simulate $e^{-iHt}$ with a PQC. As a prototype, the structure of the circuit involves all two-body interactions as previous depictions\cite{supp}.
With the ancilla-assisted PQC(Fig.\ref{APQC_demo}) employed, the objective function for both strategies can be expressed as 
\begin{eqnarray}
 f(\bm{\theta})=tr(\rho_t\Omega_{3}).
\end{eqnarray}
where $\bm{\theta}=\bm{\alpha},\bm{\beta}$ are for \emph{strategy 1} and \emph{2}. The target state has following forms,
\begin{eqnarray}
   &\rho_t=U_D \cdot e^{-iHt} \otimes I_A\cdot U_E\left(\Omega_0\right) \quad &\mbox{(\emph{strategy 1})}, \nonumber \\
   &\rho_t=\Omega_0 \quad \quad &\mbox{(\emph{strategy 2})}. \nonumber 
\end{eqnarray}
And additionally, $\Omega_3= U_D \cdot U\otimes I_A\cdot U_E\left(\Omega_0\right)$, where  
\begin{eqnarray}
   &U=U_1(\bm{\alpha})=\prod_{i=1}^N e^{-it \alpha_i \rho_i} \quad & \mbox{(\emph{strategy 1})},\nonumber \\
   &U=U_2(\bm{\beta})=U_i^{-n_c}U_{i+1}(\bm{\beta}) \quad &\mbox{(\emph{strategy 2})}. \nonumber 
\end{eqnarray}
Therefore, the target is to find the appropriate parameters $\bm{\theta}$, which maximize the objective function.
As objective function can be evaluated either with classical simulation by $\emph{strategy 1}$ or measured at the end of circuit by $\emph{strategy 2}$, the parameters can be updated according to conventional methods. 

To support the feasibility of both strategies, numerical experiments are investigated, including applications of Hamiltonian simulation and density matrix exponentiation.
In our framework, $\rho$ is taken as (a) bell state; (b) GHZ state and (c) Hamiltonian of a liquid NMR sample, Crotonic acid, which is specified in the Appendix. 
And, we numerically study the evolutions of $\rho$ with a $2\log{d}$-qubit register employed for a $\mathcal{D}(\rho)=d$  problem, where $d=2^2$, $2^3$ and $2^4$.

In this prototypical simulation, the programme is conducted according to the APQC with assumptions that all one or two quantum gates are accurate. For generating the PQC, a simpler structure is employed, which does not require all two-body interactions. the protocol is as follows. First, $\rho$ is expressed as a summation of Pauli terms, which is natural for Hamiltonian simulation. Then, only two-body Pauli terms are considered and divided into different layers of the circuit. Finally, local quantum gates are placed in the middle of layers. As a result, three circuits with the depths of $3$, $4$ and $5$ are employed for the simulation. The numbers of parameters to be optimized are thus  $12$, $24$ and $40$, which converge to $\mathcal{O}(m\times n)$, where $m$ is the depth and $n$ is the size of circuit. As the comparison,  with $\epsilon$ tolerant error, $\mathcal{O}(3/\epsilon)$, $\mathcal{O}(4/\epsilon)$ and $\mathcal{O}(5/\epsilon)$ depth circuits are required for simulation using lie-product decomposition. For density matrix exponentiation, $\mathcal{O}(1/\epsilon)$ copies of density matrix are required as with $\mathcal{O}(1/\epsilon)$-depth circuit.
Thus, the objective functions can be estimated by measuring the simulated circuit. 

As for finding optimal parameters configuration, a gradient-based method is employed as our training method, which is supposed to be improved for complicated cases in other work.
To begin with this training, random numbers by a single uniformly distribution are generated and assigned as the value for initial parameters.
During the training process, the approximation of the partial derivatives can be estimated by following symmetric difference quotient
\begin{eqnarray}
   \partial_j f =\frac{ f(\bm{\theta}+\Delta \theta_j)-f(\bm{\theta}-\Delta \theta_j)}{2 \Delta \theta_j},
\end{eqnarray}
which can be realized by repeatedly running the training circuit with a small perturbation on the $\bm{\theta}$, where $\Delta \theta_j$ is naively chosen as $0.01$ in our simulation.
Therefore, the optimal configuration of the parameters can be obtained by repeating  
\begin{eqnarray}
   \bm{\theta}=\bm{\theta} + \eta \bm{\nabla} f(\bm{\theta}),
   \label{simu_iter}
\end{eqnarray}
where $\bm{\nabla} f(\bm{x})=( \partial_1 f, \partial_2 f,...)$ and the learning rate, $\eta$, is fixed at $0.02$. For simulating \emph{strategy 1} and \emph{strategy 2}, the maximum numbers for repeating Eq.(\ref{simu_iter}), i.e., iterations are set as $300$ and $350$, respectively.

When collecting results, the strategies are different for $\emph{1}$ and $\emph{2}$. 
For \emph{strategy 1}, once the optimization completed, the parameters is supposed to be optimal for simulating a $t$-time interval evolution. Otherwise, we need to rerun the simulation and repeating (\ref{simu_iter}) for $300$ times at once until the objective function is satisfied with the optimized threshold. 
For \emph{strategy 2}, optimizing APQC is not once and for all. After the $i$-th repetition of training APQC with (\ref{simu_iter}) for $350$ iterations, only $U_i^{-n_c}U_{i+1}(\bm{\beta})$ is trained and is optimized to be identity, where $n_c$ is chosen as $2$. Therefore, $\mathcal{O}(\log(t/\Delta t))$ repetitions are required. In our simulation, $\Delta t/t=1/2^{10}$ and the base number, $2$, determines the compression efficiency of the circuit. 
\begin{figure}[h!]
   \centering
   \includegraphics[width=0.8\textwidth]{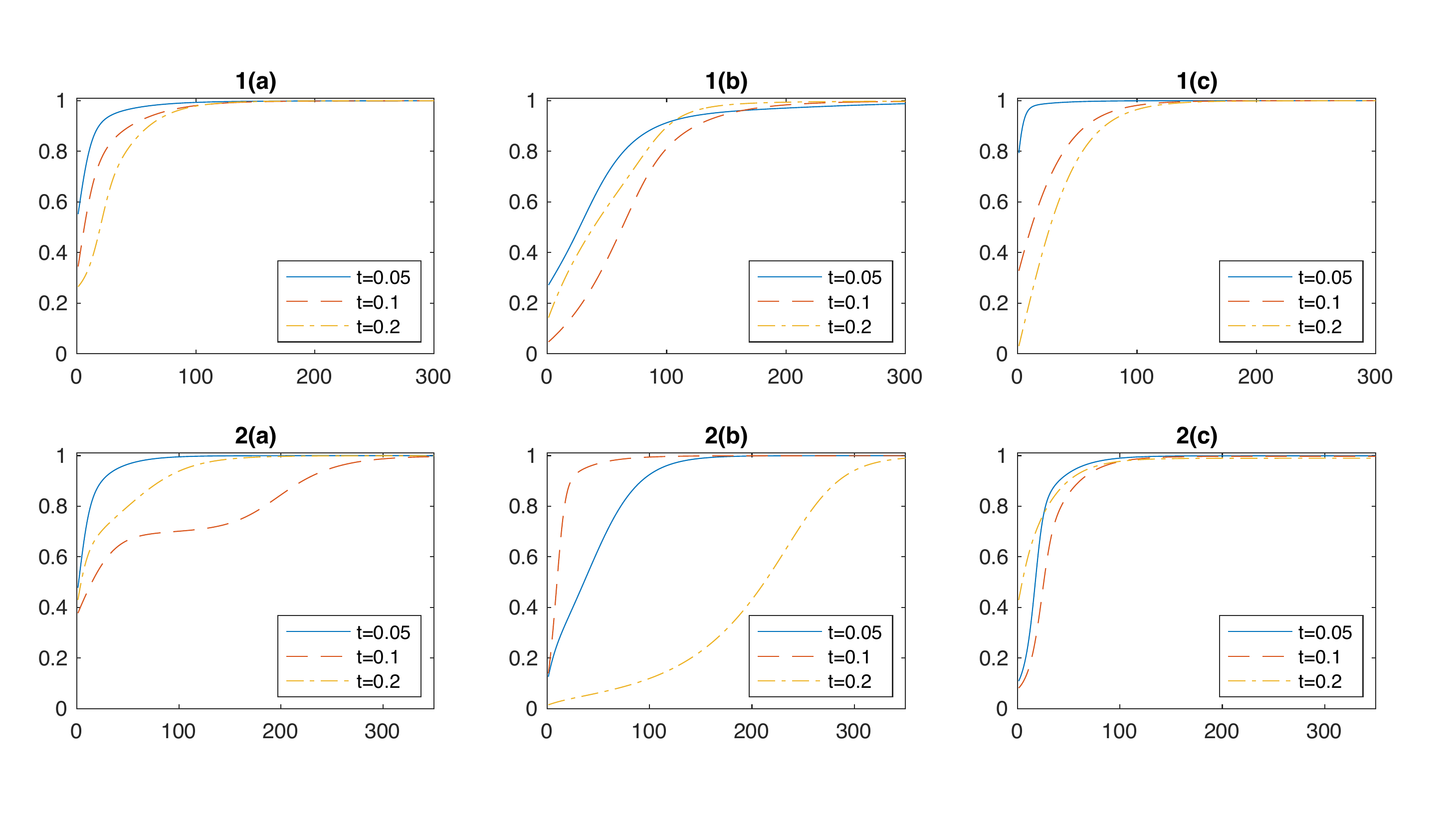}
   \caption{ Results of \emph{algorithm 1} and \emph{algorithm 2} for simulating the evolutions by $\rho$ for a period of $t=0.05$, $0.1$ and $0.2$. The fidelities(vertical axis) vary with the iterations(horizon axis). $\rho$ is chosen as (a) bell state; (b) GHZ state and (c) Hamiltonian of a molecular Crotonic acid system. } 
   \label{ResultSimu}
\end{figure}

Fig.\ref{ResultSimu} shows the feasibility of the algorithms on results which converge to the target matrix exponentiations.  
(a) (b) and (c) simulated the evolutions of bell state, GHZ state and Hamiltonian of Crotonic acid, respectively. For the each case, the time scale with $t=0.05$, $0.1$ and $0.2$ for evolution are considered. 
With the accurate local quantum gate assumption, the simulation by both \emph{ algorithms 1} and \emph{algorithm 2} generates the optimal configurations for parameters in APQC, which both approach the evolution of either density matrix or Hamiltonian as the number of iterations gets larger.


\section{Conclusion}

As the NISQ era comes in and would lasts decades, finding a near-future algorithm which exploits the power of NISQ hardwares becomes more and more important. 
Simulating matrix exponentiation, which is an essential module for many quantum algorithms, such as density matrix exponentiation for quantum principal component analysis and its variants, and Hamiltonian simulation for quantum simulations. Its implementation usually rely on a deep quantum circuit, which is intractable with NISQ devices.

In this paper, an NISQ heuristic framework is proposed for simulating $e^{-i \rho t}$ and its variants, which employs parameterized quantum circuit. To construct these PQCs, two strategies are given. \emph{strategy 1} is with the classical optimal control theory and \emph{strategy 2} is with the hybrid quantum-classical strategy. 
In order to train a useful PQC,  $\mathcal{O}(4^n\times r)$ operations are required for each iteration of \emph{strategy 1}, where $n$ is the size of quantum circuit and $r$ is the number of total iterations. Accordingly, as the size of system gets larger, the computation resources would be put into their limits.
This could be addressed by \emph{the second strategy} which is hybridized with  quantum agent. As for the training process of \emph{the second strategy}, the time complexity is $\mathcal{O}(m\times n\times r\times \log{(t/\Delta t)})$, with $2n$ being the size of circuit and $m$ being the depth of PQC. $e^{-i\rho \Delta t}$ is the initial circuit for simulation, and $\log{(t/\Delta t)}$ is number of steps to approach $e^{i\rho t}$. 
Although it is inefficient to obtain the target PQC using \emph{the first strategy}, the error is $\mathcal{O}(\epsilon_o)$, where only optimized threshold is considered . While the deviations of \emph{the second strategy} should contain two part, $\mathcal{O}((t^2/\epsilon_t)\epsilon_o)+\mathcal{O}(\epsilon_t)$, where $\epsilon_t$ is the trotter error and $\epsilon_o$ is the optimized threshold. The strategy would perform better if $\epsilon_o$ is sufficiently small.
In addition, the numerical experiments are studied with $\rho$ being considered as bell state, GHZ state and Hamiltonian of Crotonic acid, which demonstrate the feasibility of both strategies.
Remarkably, APQC is proposed in this article, which is a generalization of linear combination unitaries(LCU) and PQC. 
Though it is employed to characterize a quantum process and compress quantum circuits in this work, it is essential as an important part for variant quantum algorithms, which are important constitutions for NISQ algorithms.

Many optimization problems are in fact NP-hard problems. The strategy which are essentially optimization-based cannot get over them, too.
In addition, focusing on the gradient-based methods, they cannot avoid the local optimum problems, especially when the feasible region is complicated. Therefore, the proposed strategies would fail in finding a global optimum with a bad initial guess, just as the classical cases. 
Therefore, although the numerical results seems good without considering the initialization, a good initial guess is still crucial to decrease the possibility of failure, especially for dealing with a larger problem. 
A potential method to have a good guess is to initialize the parameters via naively applying trotter decompositions, whose infidelity is determined by the commutators generated by their decompositions. Additionally, to avoid the local optimum, we can add a random disturbance to parameters when the cost function is converging.  
Accordingly, the optimization problem is essentially difficult, not only faced with this heuristic strategy but for all optimization-based methods. Investigations on the optimization method itself, not only benefit the exploitation of near-future quantum devices but also for more optimization modern technologies\cite{ruder2016overview,gill2019practical}. 

As simulating hermitian matrix exponentiation is important for quantum information processing, our method contribute to its experimental implementation for a larger scale machine.
In spite of the problems of optimization itself, the framework with corresponding strategies using an affordable optimization method, paves a way for applications on NISQ devices of matrix exponentiation and thus contributes to the field of NISQ algorithm. In addition, APQC extends the abilities of PQC, generalizing a state-based algorithm and compressing a quantum circuit in this article, which is promising to be an important subroutine in near future and provides a new way to exploit the power of NISQ devices.

\textbf{Data availability}
All data for the figure and table are available on request. All other data about experiments are available upon reasonable request.

\textbf{Code availability}
Code used for generate the quantum circuit and implementing the experiment is available on reasonable request.

\textbf{Acknowledgements}
This research was supported by National Basic Research Program of China.
K.L. acknowledge the National Natural Science Foundation of China under grant No. 11905111.

\textbf{Competing interests}
The authors declare no competing interests.


\newpage
\section{Supplementary materials for "Towards a NISQ algorithm to simulate Matrix exponentiation"}
\subsection{A. Derivations in \emph{the second strategy}}

The objective function of \emph{the second strategy} is defined as,
\begin{eqnarray}
   f(\bm{\beta})=\langle e^{-i\Delta t\rho}\sigma e^{  i\Delta t\rho}\prod_i e^{-i\Delta t\beta_i\rho_i} \sigma \prod_i e^{i\Delta t\beta_i\rho_i}\rangle,
\end{eqnarray}
where $\sigma$ is an arbitrary density matrix and the operators in $\langle \cdot \rangle$ satisfy the permutation equality. The target is to simulate $e^{-i\Delta t\rho}$ with $\prod_i e^{-i\Delta t\beta_i\rho_i}$.  
The matrix exponentiation are expanded with the Taylor series on $\Delta t$,
\begin{eqnarray}
   e^{-i\Delta t\rho}&=&\sum_{k=0}^{\infty}{(-i\Delta t\rho)^k/k!}, \nonumber \\ 
   \prod_{i=1}^N e^{-i\Delta t \beta_i \rho_i}&=&\prod_{i=1}^N \sum_{k=0}^{\infty}{(-i\Delta t\beta_i \rho_i )^k/k!}.
\end{eqnarray}
With the following substitutions
\begin{eqnarray}
   H&=&-i\Delta t\rho,\nonumber \\
   M_i&=&-i\Delta t\beta_i \rho_i,
\end{eqnarray}
Ingredients in $\langle \cdot \rangle$ are rewritten as,
\begin{eqnarray}
   e^{i\Delta t\rho}\times \prod_{i=N}^1 e^{-i\Delta t \beta_i \rho_i}
   &&=\Scale[1]{\left(\frac{(-H)^0}{0!}+\frac{(-H)^2}{1!}+\frac{(-H)^3}{3!}+\cdots\right)\times\left(\frac{M_N^0}{0!}+\frac{M_N^1}{1!}+\frac{M_N^2}{2!}+\cdots\right)...
\left(\frac{M_1^0}{0!}+\frac{M_1^1}{1!}+\frac{M_1^2}{2!}+\cdots\right)} \nonumber \\
&&=\sum_{k=0}^{\infty}\sum_{\substack{j_i,j_{\rho}\geq 0,\\\sum_i{j_i}+j_{\rho}=k}}\frac{(-H)^{j_{\rho}}}{j_{\rho}!}
\prod_{i=N}^1 \frac{M_i^{j_i}}{j_i!},\nonumber \\ 
\nonumber \\ \nonumber \\
\prod_{i=1}^N e^{i\Delta t \beta_i \rho_i}\times e^{-i\Delta t\rho}&&=\Scale[1]{\left(\frac{(-M_1)^0}{0!}+\frac{(-M_1)^1}{1!}+\frac{(-M_1)^2}{2!}+\cdots\right)
   \cdots \left(\frac{(-M_N)^0}{0!}+\frac{(-M_N)^1}{1!}+\frac{(-M_N)^2}{2!}+\cdots\right)
 \times \left(\frac{H^0}{0!}+\frac{H^2}{1!}+\frac{H^3}{3!}+\cdots\right)}\nonumber \\
&&=\sum_{k=0}^{\infty}\sum_{\substack{j_i,j_{\rho}\geq 0,\\\sum_i{j_i}+j_{\rho}=k}}\prod_{i=1}^N \frac{(-M_i)^{j_i}}{j_i!}   \frac{H^{j_{\rho}}}{j_{\rho}!}. \nonumber 
\end{eqnarray} 
Therefore, the objective function with no approximation is 
\begin{eqnarray}
   f=\langle \psi| \sum_{k=0}^{\infty}\sum_{\substack{j_i,j_{\rho}\geq 0,\\\sum_i{j_i}+j_{\rho}=k}}\prod_{i=1}^N \frac{(-M_i)^{j_i}}{j_i!}   \frac{H^{j_{\rho}}}{j_{\rho}!}
    |\psi\rangle\langle \psi| 
    \sum_{k=0}^{\infty}\sum_{\substack{j_i,j_{\rho}\geq 0,\\\sum_i{j_i}+j_{\rho}=k}}\frac{(-H)^{j_{\rho}}}{j_{\rho}!}
    \prod_{i=N}^1 \frac{M_i^{j_i}}{j_i!}
   |\psi\rangle.
\end{eqnarray}
It is a taylor-like polynomial with respect to $\Delta t$. As $\Delta t$ can be a sufficiently small quantity, only lower order terms on $\Delta t$ are considered, such as $k=0,1,2$. For the objective function $f$, all terms which is linear with or higher than $\mathcal{O}(\Delta t^2)$ are ignored. Therefore, $f$ can be simplified as,
\begin{eqnarray}
   (1+<\frac{H^2}{2!}>-<H\sum_i M_i>+<\sum_i \frac{M_i^2}{2!}>+<\sum_{i>j}M_iM_j>)^2 -(<H>-<\sum M_i>)^2.
\end{eqnarray}
And $1-f$ is the inaccuracy and $\langle\cdot\rangle$ is the expectation value on $\ket{\psi}$, which is 
\begin{eqnarray}
   (<H>-<\sum M_i>)^2-2<\frac{H^2}{2!}>+2<H\sum_i M_i>-2<\sum_i \frac{M_i^2}{2!}>-2<\sum_{i>j}M_iM_j>
\end{eqnarray}
Thus,
\begin{eqnarray}
   \sum M_i=H \label{Trotter-decom3}
\end{eqnarray}
is the requirement for eliminating above lowest order error. In the original representation,
$\beta_i$ satisfy following correspondence 
\begin{eqnarray}
   \sum \beta_i\rho_i=\rho.\label{Trotter-decom4}
\end{eqnarray}
which is also the result by Trotter-suzuki decomposition.

\subsection{B. Method to generate a $N$-layer PQC}
In this section, a method to generate a $N$-layer PQC with all two-body interactions is introduced,where $N$ is the size of the system.
$H_2(i,j)$ are assumed as the products of Pauli matrices, where $H_2(i,j)=\sigma_k\otimes\sigma_l$ are on both $i$-th and $j$-th qubits and $i,j=1,...,N$ and $k,l=x,y,z$.
All of them can be presented in a $N\times N$ triangle parameter matrix $P$,
\begin{eqnarray}
   P=\begin{bmatrix}
       1 & H_2(1,2)& H_2(1,3) & H_2(1,4)& H_2(1,5) &,\cdots, & H_2(1,N)\\
              &  2 & H_2(2,3) & H_2(2,4)& H_2(2,5) &,\cdots, & H_2(2,N)\\
              &        &    3 &H_2(3,4) & H_2(3,5) &,\cdots, & H_2(3,N)\\
              &          &        &   4 &H_2(4,5)  &,\cdots, & H_2(3,N)\\
              &          &        &         &          &,\cdots, &         \\
              &          &        &         &          &         & H_2(N-1,N)\\
              &          &        &         &          &         & N\\
   \end{bmatrix},
\end{eqnarray}
where $P_{ii}=i$, $P_{ij}=H_2(i,j)$ for $i<j$ and $P_{ij}=0$ for $i> j$. Thus, the non-zeros elements in $P$ can be divided into $N$ group with the following strategy(\ref{Group}), where $N$ is supposed to be odd.
\begin{algorithm}[H]
   \caption{Algorithm on grouping the Pauli words}
   \label{Group}
   \begin{spacing}{1.5}
    \begin{algorithmic}
    \State \textbf{Input:} Parameter matrix $P$, $N$ null sets, and a $N\times N$ label matrix $L$, where $L_{mn}=1$ for arbitrary $m,n$.
    \State \textbf{Output:} $N$ sets of Pauli words:$\mathbb{S}_1,\mathbb{S}_2, ..., \mathbb{S}_N$
    \For{$i=1,...N$}
    \For{$j=i,...,N$}
    \If{$i=j\cap L_{ij}=1$}
    \State $P_{ij}\in\mathbb{S}_i$, $L_{ij}=0$.
    \Else
    \For{$m=1,...,N$}
    \If{$L_{mi}=1\cap L_{mj}=1$}
    \State $P_{ij}\in\mathbb{S}_m$, $L_{mi}=0$ ,$L_{mj}=0$
    \EndIf
    \EndFor
    \EndIf
    \EndFor
    \EndFor
    \end{algorithmic}
   \end{spacing}
 \end{algorithm}
With the complexity of $\mathcal{O}(N^3)$ operations, Those Pauli words are divided into $N$ group, where the Pauli words in each group commute with each other. If $N$ is even, the strategy can be easily generalized. Therefore, via a matrix exponentiation formation multiplied by tunable parameters, they can be arranged into a parameterized circuit, with a depth of $N$, being linear with the size of the system. If the combinations of different type Pauli operators are considered, $\mathcal{O}(N)$ is the depth. Therefore, this algorithm provided a depth of $\mathcal{O}(N)$ parameterized circuit with all two-body interactions considered.

\subsection{C. Expressibility of the PQC}

A parameterized quantum circuit with excellent expression is more likely to represent a target state. To measure the expressive power of a circuit, expressibility is defined in recent work as a distance of two state distributions, characterized with $||\cdot||$, the Hilbert-Schmidt norm.
\begin{eqnarray}
   A=||\int_{Haar} (\ket{\psi}\bra{\psi}) ^{\otimes t} d\psi - \int_{\bm{\theta}} (\ket{\psi_{\bm{\theta}}}\bra{\psi_{\bm{\theta}}})^{\otimes t} d\psi_{\bm{\theta}} ||,
\end{eqnarray}
where $\ket{\psi}_{\bm{\theta}}$ are outputs of a PQC with randomized configuration $\bm{\theta}$ and $\ket{\psi}$ are from a distribution according to the Haar measure. It is a generalization from the investigation of pseudorandom circuit. Thus, a highly expressible circuit would produce a small $A$, with $A=0$ corresponding to being maximally expressive, i.e., generating a state distribution to the Haar measure.

In order to explicitly estimate the expressibility with a discrete simulated results, the Kullback-Leibler (KL) divergence, i.e., relative entropy is employed, which measures the difference between one probability distribution and a reference probability distribution. It is denoted as $Expr$,
\begin{eqnarray}
   Expr=D_{KL}(P_{PQC}(f|\bm{\theta},\bm{\theta'})|P_{Haar}(f)),
\end{eqnarray}
where $P_{\mbox{PQC}}(f|\bm{\theta},\bm{\theta'})$ is the probability distribution of $f$=$|\bra{\psi_{\bm{\theta}}}\psi_{\bm{\theta'}}\rangle|^2$, outputs of a PQC with random $\bm{\theta}$ and $\bm{\theta'}$. For the reference probability distribution, i.e. to Haar measure, $P_{Haar}(f)=(N-1)(1-f)^{N-2}$ is analytical, with $N$ being size of dimension. 
$Expr$ scores the similarity to the distribution of Haar as a PQC with a lower value of KL divergence is a more expressible circuit.

\begin{figure}[!h]
   \centerline{
\resizebox{1.\textwidth}{!}{
   \begin{tikzpicture}[thick]
   \ctikzset{scale=1.65}
   \tikzstyle{every node}=[font=\normalsize,scale=1]
   \tikzstyle{operator} = [draw,shape=rectangle,fill=white,minimum size=2em] 
     \tikzstyle{operator2} = [draw,shape=rectangle,fill=white,minimum size=2em] 
     \tikzstyle{operator3} = [draw,fill=white,minimum size=4em] 
     \tikzstyle{operator4} = [draw,dashed,shape=rectangle, minimum width=7.5em, minimum height=9.5em] 
     \tikzstyle{operator5} = [draw,dashed,shape=rectangle, minimum width=12em, minimum height=9.5em] 
     \tikzstyle{operator6} = [draw,dashed,shape=rectangle, minimum width=38em, minimum height=9.5em] 
     \tikzstyle{phase} = [fill,shape=circle,minimum size=3pt,inner sep=0pt]
     \tikzstyle{surround} = [fill=blue!10,thick,draw=black,rounded corners=2mm]
     \tikzstyle{ellipsis} = [fill,shape=circle,minimum size=2pt,inner sep=0pt]
     \tikzstyle{ellipsis2} = [fill,shape=circle,minimum size=0.5pt,inner sep=0pt]
     \node at (1-0.1,-1) (q1) {$\ket{0}$};
     \node (end3) at (3,-1) {} edge [-] (q1);
     \node at (1-0.1,-1.5) (q2) {$\ket{0}$};
     \node (end3) at (3,-1.5) {} edge [-] (q2);
     \node at (1-0.1,-2) (q3) {$\ket{0}$};
     \node (end3) at (3,-2) {} edge [-] (q3);
     \node at (1-0.1,-2.5) (q4) {$\ket{0}$};
     \node (end3) at (3,-2.5) {} edge [-] (q4);
     \node[operator] (op22) at (1.5,-1) {$R_X$} ;
     \node[operator] (op22) at (1.5,-1.5) {$R_X$} ;
     \node[operator] (op22) at (1.5,-2) {$R_X$} ;
     \node[operator] (op22) at (1.5,-2.5) {$R_X$} ;
     \node[operator] (op22) at (2.5,-1) {$R_Z$} ;
     \node[operator] (op22) at (2.5,-1.5) {$R_Z$} ;
     \node[operator] (op22) at (2.5,-2) {$R_Z$} ;
     \node[operator] (op22) at (2.5,-2.5) {$R_Z$} ;
     \node[operator4] (op22) at (2,-1.75) {} ;
     \node at (2,-3) (rho) {Circuit 1};
     \node at (3.5-0.1,-1) (q1) {$\ket{0}$};
     \node (end3) at (6.5,-1) {} edge [-] (q1);
     \node at (3.5-0.1,-1.5) (q2) {$\ket{0}$};
     \node (end3) at (6.5,-1.5) {} edge [-] (q2);
     \node at (3.5-0.1,-2) (q3) {$\ket{0}$};
     \node (end3) at (6.5,-2) {} edge [-] (q3);
     \node at (3.5-0.1,-2.5) (q4) {$\ket{0}$};
     \node (end3) at (6.5,-2.5) {} edge [-] (q4);
     \node[operator] (op22) at (4,-1) {$H$} ;
     \node[operator] (op22) at (4,-1.5) {$H$} ;
     \node[operator] (op22) at (4,-2) {$H$} ;
     \node[operator] (op22) at (4,-2.5) {$H$} ;
     \node[operator] (op22) at (6,-1) {$R_X$} ;
     \node[operator] (op22) at (6,-1.5) {$R_X$} ;
     \node[operator] (op22) at (6,-2) {$R_X$} ;
     \node[operator] (op22) at (6,-2.5) {$R_X$} ;
     \node[phase] (ph11) at (4.5,-2) {};
     \node[phase]  (ph22) at (4.5,-2.5) {} edge [-] (ph11);
     \node[phase] (ph11) at (5,-1.5) {};
     \node[phase]  (ph22) at (5,-2) {} edge [-] (ph11);
     \node[phase] (ph11) at (5.5,-1) {};
     \node[phase]  (ph22) at (5.5,-1.5) {} edge [-] (ph11);
     \node[operator5] (op22) at (5,-1.75) {} ;
     \node at (5,-3) (rho) {Circuit 2};
     \node at (7-0.1,-1) (q1) {$\ket{0}$};
     \node (end3) at (15.5,-1) {} edge [-] (q1);
     \node at (7-0.1,-1.5) (q2) {$\ket{0}$};
     \node (end3) at (15.5,-1.5) {} edge [-] (q2);
     \node at (7-0.1,-2) (q3) {$\ket{0}$};
     \node (end3) at (15.5,-2) {} edge [-] (q3);
     \node at (7-0.1,-2.5) (q4) {$\ket{0}$};
     \node (end3) at (15.5,-2.5) {} edge [-] (q4);
     \node[operator] (op22) at (7.5,-1) {$R_X$} ;
     \node[operator] (op22) at (7.5,-1.5) {$R_X$} ;
     \node[operator] (op22) at (7.5,-2) {$R_X$} ;
     \node[operator] (op22) at (7.5,-2.5) {$R_X$} ;
     \node[operator] (op22) at (8,-1) {$R_Z$} ;
     \node[operator] (op22) at (8,-1.5) {$R_Z$} ;
     \node[operator] (op22) at (8,-2) {$R_Z$} ;
     \node[operator] (op22) at (8,-2.5) {$R_Z$} ;
     \node[operator] (ph11) at (8.5,-2) {$R_X$};
     \node[phase]  (ph22) at (8.5,-2.5) {} edge [-] (ph11);
     \node[operator] (ph11) at (9,-1.5) {$R_X$};
     \node[phase]  (ph22) at (9,-2.5) {} edge [-] (ph11);
     \node[operator] (ph11) at (9.5,-1) {$R_X$};
     \node[phase]  (ph22) at (9.5,-2.5) {} edge [-] (ph11);
     \node[operator] (ph11) at (10,-2.5) {$R_X$};
     \node[phase]  (ph22) at (10,-2) {} edge [-] (ph11);
     \node[operator] (ph11) at (10.5,-1.5) {$R_X$};
     \node[phase]  (ph22) at (10.5,-2) {} edge [-] (ph11);
     \node[operator] (ph11) at (11,-1) {$R_X$};
     \node[phase]  (ph22) at (11,-2) {} edge [-] (ph11);
     \node[operator] (ph11) at (11.5,-2.5) {$R_X$};
     \node[phase]  (ph22) at (11.5,-1.5) {} edge [-] (ph11);
     \node[operator] (ph11) at (12,-2) {$R_X$};
     \node[phase]  (ph22) at (12,-1.5) {} edge [-] (ph11);
     \node[operator] (ph11) at (12.5,-1) {$R_X$};
     \node[phase]  (ph22) at (12.5,-1.5) {} edge [-] (ph11);
     \node[operator] (ph11) at (13,-2.5) {$R_X$};
     \node[phase]  (ph22) at (13,-1) {} edge [-] (ph11);
     \node[operator] (ph11) at (13.5,-2) {$R_X$};
     \node[phase]  (ph22) at (13.5,-1) {} edge [-] (ph11);
     \node[operator] (ph11) at (14,-1.5) {$R_X$};
     \node[phase]  (ph22) at (14,-1) {} edge [-] (ph11);
     \node[operator] (op22) at (14.5,-1) {$R_X$} ;
     \node[operator] (op22) at (14.5,-1.5) {$R_X$} ;
     \node[operator] (op22) at (14.5,-2) {$R_X$} ;
     \node[operator] (op22) at (14.5,-2.5) {$R_X$} ;
     \node[operator] (op22) at (15,-1) {$R_Z$} ;
     \node[operator] (op22) at (15,-1.5) {$R_Z$} ;
     \node[operator] (op22) at (15,-2) {$R_Z$} ;
     \node[operator] (op22) at (15,-2.5) {$R_Z$} ;
     \node[operator6] (op22) at (11.25,-1.75) {} ;
     \node at (11.25,-3) (rho) {Circuit 3};
   \end{tikzpicture} } }
   \caption{A set of circuit templates considered in the study, each labeled with a circuit ID. The dashed box indicates a single circuit layer, denoted by $L$ in the text, that can be repeated.}   
\label{PQC_ID} 
\end{figure}
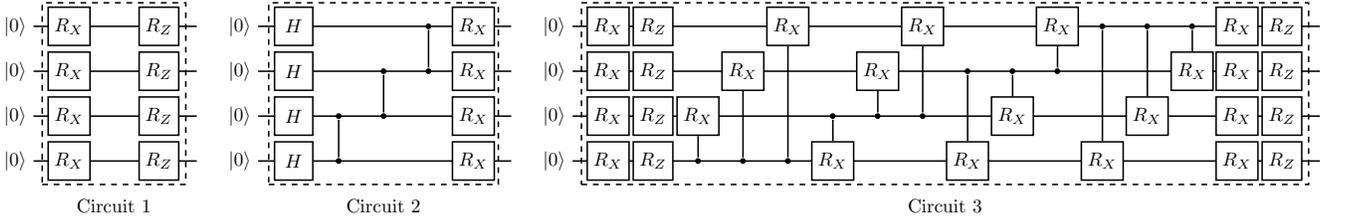

\begin{figure}[!h]
   \centerline{
   \begin{tikzpicture}[thick]
   \ctikzset{scale=1.5}
   \tikzstyle{every node}=[font=\normalsize,scale=1]
   \tikzstyle{operator} = [draw,fill=white,minimum size=1em] 
     \tikzstyle{operator2} = [draw,fill=white,minimum size=2em] 
     \tikzstyle{operator3} = [draw,fill=white,minimum size=4em] 
     \tikzstyle{operator4} = [draw,shape=rectangle, fill=white, minimum width=1em, minimum height=10em] 
     \tikzstyle{phase} = [fill,shape=circle,minimum size=3pt,inner sep=0pt]
     \tikzstyle{surround} = [fill=blue!10,thick,draw=black,rounded corners=2mm]
     \tikzstyle{ellipsis} = [fill,shape=circle,minimum size=2pt,inner sep=0pt]
     \tikzstyle{ellipsis2} = [fill,shape=circle,minimum size=0.5pt,inner sep=0pt]
     \node at (3.5,-2) (q2) {$A$};
     \node (end3) at (7,-2) {} edge [-] (q2);
     \node at (3.5,-3) (q3) {$S$};
     \node (end3) at (7,-3) {} edge [-] (q3);
     \node at (3.8,-3) (hy111) {$/$};
     \node[operator2] (op22) at (4.5,-2) {$H$} ;
     \node[operator2] (op22) at (4.5,-3) {$U_{\bm{\theta}}$} ;
     \node[phase] (ph11) at (5.5,-2) {};
     \node[operator2]  (op22) at (5.5,-3) {$U_{\bm{\theta'}}\cdot U_{\bm{\theta}}^{\dagger}$} edge [-] (ph11);
     \draw [dashed] (6.25,-1.7) -- (6.25,-3.3);
     \node at (6.25,-3.5) (rho2) {$\rho$};
     \node[meter] (op22) at (7,-2) {} ;
     \node at (7.6,-2) (q3) {$\sigma_x/\sigma_y$};
   \end{tikzpicture} } 
   \caption{Circuit for measurement of the expressivity of the PQC. It costs one extra qubit as ancillary. $\bm{\theta}$ and $\bm{\theta'}$ are two random parameter configurations for PQCs $U_{\bm{\theta}}$ and $U_{\bm{\theta'}}$, $H$ is a one-qubit Hadamard gate. }   
\label{expressivity_extract} 
\end{figure}
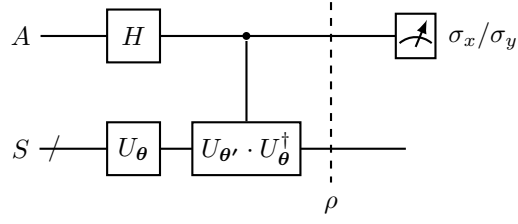

Accordingly, numerical experiments to compute and compare the expressibility in the strategy with three other typical PQCs  are studied, which are shown in Fig.\ref{PQC_ID}. They are all 4 qubit circuit.
As with the original work, to construct a distribution of $P_{PQC}$ with histogram, a bin size is set as $75$, and $5,000$ $f$ are measured.
For each type of PQCs except for the one in our strategy, the instances where the number of layers are investigated with up to 5. 
Compared with $10,000$ times repeatedly running the circuits and state tomography,  $f$ can be measured directly with the setup proposed in Fig.\ref{expressivity_extract} for $5,000$ time and without tomography. Only 1-qubit is added as the ancillary system $A$, at the end of circuit, the system is evolved as
\begin{eqnarray}
   \rho=\ket{0}\bra{0}\otimes \ket{\psi_{\bm{\theta}}}\bra{\psi_{\bm{\theta}}}+\ket{1}\bra{0}\otimes \ket{\psi_{\bm{\theta'}}}\bra{\psi_{\bm{\theta}}}+\ket{0}\bra{1}\otimes \ket{\psi_{\bm{\theta}}}\bra{\psi_{\bm{\theta'}}}+\ket{1}\bra{1}\otimes \ket{\psi_{\bm{\theta'}}}\bra{\psi_{\bm{\theta'}}}.
\end{eqnarray}

Via the measurements on ancillary system, $f$ can be obtained with the real and imaginary part of $\langle \bra{\psi_{\bm{\theta}}}\psi_{\bm{\theta'}}\rangle \rangle$, which are extracted as 
\begin{eqnarray}
   \mbox{Re}&=&tr(\sigma_x \otimes I \cdot \rho )=\frac{1}{2}(\langle\psi_{\bm{\theta}}|\psi_{\bm{\theta'}}\rangle+\langle\psi_{\bm{\theta'}}|\psi_{\bm{\theta}}\rangle)\nonumber \\
   \mbox{Im}&=&tr(\sigma_y\otimes I \cdot \rho )= \frac{-i}{2}(\langle\psi_{\bm{\theta}}|\psi_{\bm{\theta'}}\rangle-\langle\psi_{\bm{\theta'}}|\psi_{\bm{\theta}}\rangle).
\end{eqnarray}
Results are shown in the figure of the main article.
For three types of PQCs which are denoted in circle, diamond and square, the instances are considered where the circuit, which is as a unit layer, are repeated with up to 5. 

In general, repeating a circuit layer would increase the expressibility.
As there is no entanglement gate in circuit 1, this argument does not hold. For circuit 2 and 3, even if a single layer for circuit 3 have pretty good expression, this argument holds. From the result, there are convergences for the metric of expressibility. For the argument wether more numbers of layers can help the expression, at least, we can not obtain more information from this metric. 
Additionally, the PQC generated in our strategy, which is labeled as cyan cross, keeps a same structure. It have a similar performance with respect to the repeated circuits 2 and 3.
As a small value of expressibility imply an excellent expression, the generated circuits in the strategy performs no worse than other typical circuit.

\subsection{D. Objective function for Hamiltonian Simulation}

In this section, the explicit expression of the objective functions for Hamiltonian simulation is derived, as well as the step-by-step illustrations of the circuit to implement the Hamiltonian simulation.
The intermediate state and quantum operations of the circuit are depicted as follows.

\begin{eqnarray}
   \Omega_0&=&\ket{\psi_0}\bra{\psi_0},\quad \ket{\psi_0}=\ket{0}_S \ket{0}_A,\nonumber \\
   \Omega_1&=&\ket{\psi_1}\bra{\psi_1},\quad \ket{\psi_1}=U_E \ket{\psi_0},
\nonumber \\
   \Omega_2&=&\ket{\psi_2}\bra{\psi_2},\quad \ket{\psi_2}=U_{PQC} \otimes I_A \ket{\psi_1},\nonumber \\
   \Omega_3&=&\ket{\psi_3}\bra{\psi_3},\quad \ket{\psi_3}=U_D\ket{\psi_2}.
\end{eqnarray}
where 
\begin{eqnarray}
   U_E&=&I_S \otimes H_A \cdot Cz \cdot H_S \otimes H_A,\nonumber \\
   U_D&=&H_S \otimes H_A \cdot  Cz \cdot I_S \otimes H_A.
\end{eqnarray}

\begin{eqnarray}
   |\psi_1\rangle=\sum_{i=1}^{d}\ket{i}_S\otimes\ket{i}_A
\end{eqnarray}
is the bell state pairs, where the idea of Choi matrix for tomography is employed. 

With respect to $U_{PQC}$, for \emph{the first strategy}, it aims at finding optimal parameters $\bm{\alpha}=(\alpha_1,\alpha_2,...)$ to generate the Hamiltonian simulation PQC. While, for the $i$-th repetition of \emph{the second strategy}, it aims at finding appropriate parameters $\bm{\beta}=(\beta_1,\beta_2,...)$, generating a PQC which realizes repeatedly applied $U_i$. 
Thus, the different $U_{PQC}$, which are to be optimized, are shown as
\begin{eqnarray}
   &U_{PQC}=U_1(\bm{\alpha})=\prod_{i=1}^N e^{-it \alpha_i \rho_i} \quad & \mbox{(\emph{the first strategy})},\nonumber \\
   &U_{PQC}=U_2(\bm{\beta})=U_i^{-2}U_{i+1}(\bm{\beta}) \quad &\mbox{(\emph{the second strategy})}. \nonumber 
\end{eqnarray}

As for the the target states, they are also different. \emph{The first strategy} realizes a evolution by a target Hamiltonian on system $S$ while \emph{the second strategy} realizes $I_S$. Therefore, the density matrix $\rho_t$ for both strategies are
\begin{eqnarray}
   &\rho_t=U_D \cdot e^{-iHt} \otimes I_A\cdot U_E\left(\Omega_0\right) \quad &\mbox{(\emph{the fist strategy})}, \nonumber \\
   &\rho_t=\Omega_0 \quad \quad &\mbox{(\emph{the second strategy})}. \nonumber 
\end{eqnarray}
And the objective function for Hamiltonian simulation can be concluded as 
\begin{eqnarray}
   f(\bm{\theta})=tr(\rho_t\Omega_{3}),
  \end{eqnarray}
where $\bm{\theta}=\bm{\alpha},\bm{\beta}$ are for \emph{the first strategy} and \emph{the second strategy}, respectively.

\subsection{E.Information for Numerical simulation}
In this section, the specified information for $\rho$ in simulation part are given. Three cases are simulated and they are as follows.

\emph{1. Bell State} 
\begin{eqnarray}
   \ket{\psi}=\frac{1}{\sqrt{2}}(\ket{00}+\ket{11}), \quad \rho=\frac{1}{2}\left(
   \begin{matrix}
      1&0&0&1\\
      0&0&0&0\\
      0&0&0&0\\
      1&0&0&1
   \end{matrix}\right).
\end{eqnarray}

\emph{2. GHZ State} 
\begin{eqnarray}
   \ket{\psi}=\frac{1}{\sqrt{2}}(\ket{000}+\ket{111}), \quad \rho=\frac{1}{2}\left(
      \begin{matrix}
         1&0&0&0&0&0&0&1\\
         0&0&0&0&0&0&0&0\\
         0&0&0&0&0&0&0&0\\
         0&0&0&0&0&0&0&0\\
         0&0&0&0&0&0&0&0\\
         0&0&0&0&0&0&0&0\\
         0&0&0&0&0&0&0&0\\
         1&0&0&0&0&0&0&1
      \end{matrix}\right).
\end{eqnarray}

\emph{3. Hamiltonian of Crotonic Acid}

The Hamiltonian simulated in our numerical experiments is from $^{13}$C-labeled Crotonic acid dissolved in d6-acetone, which is usually employed as a four-qubit quantum system for NMR-based quantum information processing. 

\begin{figure}[htb]
\begin{center}
\includegraphics[width= 0.6\columnwidth]{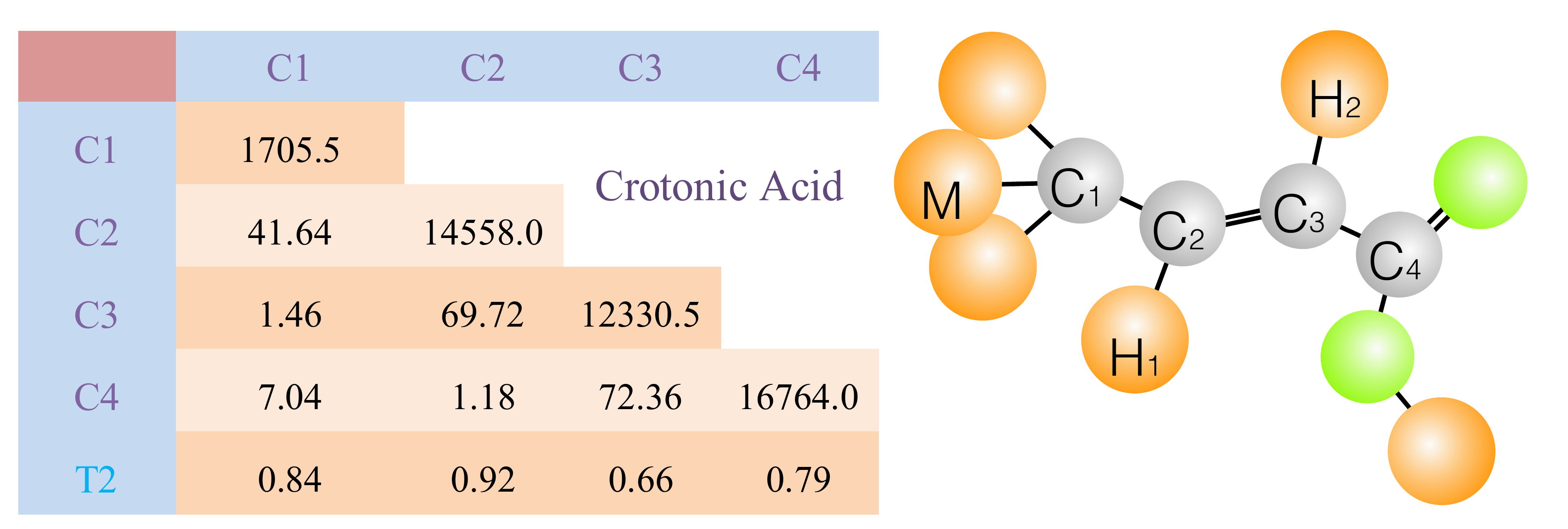}
\end{center}
\setlength{\abovecaptionskip}{-0.00cm}
\caption{ Molecular structure of $^{13}$C-labeled Crotonic acid. 
$\nu_j$ and $J_{jk}$ are the chemical shifts and J-couplings,respectively, which are listed by the diagonal and off-diagonal elements. T$_{2}$ (in Seconds) are the relaxation time which are shown at bottom.}\label{moleculeham}
\end{figure}

The structure of this molecule is shown in Fig.\ref{moleculeham} , where C$_1$ to C$_4$ are four carbon atoms, $M$ is one of three hydrogen atoms in a methyl group, H$_1$ and H$_2$ are two hydrogen atoms . When it is used as a four qubit quantum system, the hydrogen atoms are decoupled from the carbon atoms with a shaped radio frequency pulse. Under the circumstance of weak coupling, the internal Hamiltonian of this molecule can be expressed as,
\begin{eqnarray}\label{Hamiltonian}
H_{int}=\sum\limits_{j=1}^4 {\frac{1}{2} \nu _j } \sigma_z^j  + \sum\limits_{j < k}^4 {\frac{\pi}{2}} J_{jk} \sigma_z^j \sigma_z^k.
\end{eqnarray}
$\nu_j$ and $J_{jk}$ are inside Hamiltonian parameters whose values are listed in the table besides the molecule. $\nu_j$, the chemical shift, are the diagonal elements and $J_{jk}$, the J-coupling, are the off-diagonal elements. As for $T_2$, the transverse relaxation time, is not involved in our simulation as the unitary process is assumed. All parameters can be found in related paper which is under a magnetic field of  $9.4T$ at room temperature (296.5K).

\end{document}